\documentclass[acmtog,screen,nonacm]{dhara-preprint}
\makeatletter\@ACM@balancefalse\makeatother
\usepackage{booktabs}
\usepackage{tabularx}
\usepackage{array}
\usepackage{enumitem}
\usepackage{tikz}
\usetikzlibrary{arrows.meta,positioning,fit,calc}
\usepackage{microtype}
\usepackage{graphicx}
\usepackage{url}
\usepackage{fvextra}
\usepackage{placeins}
\usepackage{dblfloatfix}


\usepackage{xr-hyper}
\input{supplementary-labels.tex}
\newcommand{\dhara}{Dh\=ar\=a}

\newcommand{\NQuestions}{325}

\setcopyright{none}
\acmDOI{}
\AtBeginDocument{\renewcommand{\keywordsname}{Keywords}}
\hypersetup{pdfcreator={LaTeX}}
\renewcommand\footnotetextcopyrightpermission[1]{}

\title{\dhara: Evidence-Oriented Retrieval for Indian Knowledge Systems}

\author{Avinash Holenarasipur Nagendra}
\orcid{0009-0002-2800-5113}
\authornote{Avinash Holenarasipur Nagendra and Anand Kaniyaar Sriranga Prasad contributed equally and are joint first authors.}
\affiliation{%
  \department{Department of Heritage Science \& Technology}
  \institution{Indian Institute of Technology Hyderabad}
  \city{Sangareddy}
  \state{Telangana}
  \country{India}}
\author{Anand Kaniyaar Sriranga Prasad}
\orcid{0009-0008-1360-7334}
\authornotemark[1]
\affiliation{%
  \institution{HTSR Institute}
  \city{Bengaluru}
  \state{Karnataka}
  \country{India}}
\affiliation{%
  \department{Department of Heritage Science \& Technology}
  \institution{Indian Institute of Technology Hyderabad}
  \city{Sangareddy}
  \state{Telangana}
  \country{India}}
\author{Rahul Vishwakarma}
\orcid{0009-0001-8124-3535}
\affiliation{%
  \institution{Bheri Technology Foundation}
  \streetaddress{ITIC, Indian Institute of Technology Hyderabad}
  \city{Sangareddy}
  \state{Telangana}
  \country{India}}

\author{Trupti Gavit}
\orcid{0009-0004-9273-6273}
\affiliation{%
  \institution{Bheri Technology Foundation}
  \streetaddress{ITIC, Indian Institute of Technology Hyderabad}
  \city{Sangareddy}
  \state{Telangana}
  \country{India}}

\author{Anish Amol Tipnis}
\orcid{0009-0007-4838-9085}
\affiliation{%
  \institution{Bheri Technology Foundation}
  \streetaddress{ITIC, Indian Institute of Technology Hyderabad}
  \city{Sangareddy}
  \state{Telangana}
  \country{India}}

\author{Yashaswini Mandayam Rangayyan}
\orcid{0000-0002-6640-2390}
\affiliation{%
  \department{Department of Biomedical Engineering}
  \institution{Indian Institute of Technology Hyderabad}
  \city{Sangareddy}
  \state{Telangana}
  \country{India}}

\author{Vinod Appajanna}
\orcid{0009-0006-8024-4217}
\affiliation{%
  \institution{IBM (India)}
  \city{Bengaluru}
  \state{Karnataka}
  \country{India}}
\affiliation{%
  \department{Department of Heritage Science \& Technology}
  \institution{Indian Institute of Technology Hyderabad}
  \city{Sangareddy}
  \state{Telangana}
  \country{India}}

\author{Kousik Sarathy Sridharan}
\orcid{0000-0003-3736-1873}
\affiliation{%
  \department{Department of Heritage Science \& Technology}
  \institution{Indian Institute of Technology Hyderabad}
  \city{Sangareddy}
  \state{Telangana}
  \country{India}}
\affiliation{%
  \department{Department of Biomedical Engineering}
  \institution{Indian Institute of Technology Hyderabad}
  \city{Sangareddy}
  \state{Telangana}
  \country{India}}
\affiliation{%
  \institution{Bheri Technology Foundation}
  \streetaddress{ITIC, Indian Institute of Technology Hyderabad}
  \city{Sangareddy}
  \state{Telangana}
  \country{India}}
\author{Mohan Raghavan}
\orcid{0000-0002-9196-8193}
\affiliation{%
  \department{Department of Heritage Science \& Technology}
  \institution{Indian Institute of Technology Hyderabad}
  \city{Sangareddy}
  \state{Telangana}
  \country{India}}
\affiliation{%
  \department{Department of Biomedical Engineering}
  \institution{Indian Institute of Technology Hyderabad}
  \city{Sangareddy}
  \state{Telangana}
  \country{India}}
\affiliation{%
  \institution{Central Sanskrit University}
  \city{New Delhi}
  \country{India}}
\affiliation{%
  \institution{Bheri Technology Foundation}
  \streetaddress{ITIC, Indian Institute of Technology Hyderabad}
  \city{Sangareddy}
  \state{Telangana}
  \country{India}}
\renewcommand{\shortauthors}{Avinash et al.}

\makeatletter
\newcommand{\dharaauthoraffiliations}[2]{%
  \ifcsdef{typeset@author#1}{%
    \csletcs{dhara@typeset@author#1}{typeset@author#1}%
  }{%
    \cslet{dhara@typeset@author#1}{\@firstofone}%
  }%
  \csdef{typeset@author#1}##1{%
    \csuse{dhara@typeset@author#1}{##1}\textsuperscript{#2}}%
}
\dharaauthoraffiliations{1}{1}
\dharaauthoraffiliations{2}{2,1}
\dharaauthoraffiliations{3}{5}
\dharaauthoraffiliations{4}{5}
\dharaauthoraffiliations{5}{5}
\dharaauthoraffiliations{6}{3}
\dharaauthoraffiliations{7}{6,1}
\dharaauthoraffiliations{8}{1,3,5}
\dharaauthoraffiliations{9}{1,3,4,5}

\def\dhara@skipaffiliation#1#2{%
  \let\and\@typeset@author@line}
\patchcmd{\@mkauthors@i}
  {\noindent\addresses\@typeset@author@line}
  {\noindent\let\affiliation\dhara@skipaffiliation\addresses\@typeset@author@line}
  {}%
  {\ClassError{dhara-paper}{Cannot apply shared-affiliation layout}%
    {Check the author layout against the installed acmart class.}}
\makeatother

\authorsaddresses{%
  Affiliations: \textsuperscript{1}Department of Heritage Science \& Technology,
  Indian Institute of Technology Hyderabad, Sangareddy, Telangana, India.\newline
  \textsuperscript{2}HTSR Institute, Bengaluru, Karnataka, India.\newline
  \textsuperscript{3}Department of Biomedical Engineering,
  Indian Institute of Technology Hyderabad, Sangareddy, Telangana, India.\newline
  \textsuperscript{4}Central Sanskrit University, New Delhi, India.\newline
  \textsuperscript{5}Bheri Technology Foundation, ITIC, Indian Institute of Technology Hyderabad, Sangareddy, Telangana, India.\newline
  \textsuperscript{6}IBM (India), Bengaluru, Karnataka, India}

\begin{abstract}
Relevant evidence in Indian Knowledge Systems (IKS) collections may be distributed across passages in different parts of a work, including verses, translations, and commentaries. Locating it requires matching natural-language questions and partially remembered verse fragments to source passages. We present \dhara{}, a retrieval engine combining hybrid search, chapter- and book-level Entity Notes (EN), Fuzzy Verse Retrieval (FVR), and neural reranking. Five comparisons on the 325 questions in \dhara{} Bench examine retrieval representations against BM25, EN's contribution, reranker quality and computation time, FVR, and Gemini. BGE-M3 single-vector retrieval exceeds BM25 in nDCG@10, Recall@20, and MRR@20 across all categories. Its scores are broadly comparable to ColBERTv2, with mean retrieval times of 0.108~s and 0.962~s under the same hardware configuration and a smaller memory footprint for chunk storage on disk. Adding EN to a direct branch with a matched candidate count improves nDCG@10 in all twelve categories with Qwen and ten with BGE; the combined ranking is capped at the same candidate count. Qwen improves nDCG@10, Precision@5, and MRR@20 over BGE in every category, at approximately 6.72 times the overall estimated serial computation time. FVR increases Verse-category nDCG@10 by approximately 20 percentage points with either reranker. Gemini is evaluated after source mapping and manual review with relevance annotations visible. The Qwen pipeline with EN and FVR has higher Recall@20 than Gemini in ten categories and stronger complete-evidence retrieval in Multi-hop questions, while Gemini retains early-relevance advantages in several categories. The results inform retrieval choices according to evidence requirements and resource budgets.
\end{abstract}


\begin{document}
\raggedbottom
\maketitle
\pagestyle{empty}\thispagestyle{empty}\hypersetup{pdfcreator={LaTeX}}

\section{Introduction}
Retrieving evidence from Indian Knowledge Systems (IKS) collections often requires locating related passages across a work, including its verses, translations, and commentaries. Questions may ask about a concept, connect references to an entity across chapters, or seek the source of a partially remembered verse. Relevant passages may use different wording from the question, and some questions require evidence from several passages together.

These requirements create linked retrieval design choices. The representation of each passage affects matching, latency, and the memory footprint for passage storage. Additional retrieval paths can identify evidence through entity descriptions or verse fragments, but their contributions must be assessed alongside the reranking of the resulting source passages. Locating an early relevant passage and recovering a complete evidence group are distinct goals, so a useful comparison must consider question categories and the evidence required.

We present \dhara{}, a retrieval engine that integrates hybrid passage search, Entity Notes (EN), Fuzzy Verse Retrieval (FVR), and neural reranking. Chapter-level Entity Notes (CEN) describe entities within chapters or source units, while Book-level Entity Notes (BEN) consolidate their descriptions across chapters of a book. We hypothesize that these entity-focused descriptions can improve retrieval by providing an additional route to relevant source passages. EN records and FVR matches link to source passages for ranking. Section~\ref{sec:engine} describes their construction and integration.

The evaluation comprises two experiments using the \NQuestions{} questions, relevance annotations, and corpus passages in \dhara{} Bench~\cite{dharaBenchmark2026}. We refer to these corpus passages as \emph{source chunks}. The first compares single-vector (SV) hybrid retrieval using BGE-M3~\cite{chen2024m3} and all-MiniLM-L6-v2~\cite{allMiniLML6v2}, multivector (MV) hybrid retrieval using ColBERTv2~\cite{santhanam2022colbertv2}, and the BM25 baseline from \dhara{} Bench. The second uses the selected representation to examine four decisions: adding EN to a direct branch with a matched candidate budget, choosing BGE or Qwen reranking, enabling FVR, and comparing the combined pipeline with Gemini. Gemini serves as an LLM-based evidence retriever whose generated references are mapped to source chunks and manually reviewed before evaluation.

The paper makes three contributions:
\begin{enumerate}
\item An IKS retrieval architecture combining hybrid search, chapter- and book-level EN linked to source chunks, FVR, and neural reranking.
\item A comparison of SV, MV, and BM25 retrieval across question categories and retrieval depths, including shared and unique relevant evidence, latency, and memory footprint for chunk storage.
\item A component evaluation matching the merged candidate count to the direct-retrieval budget, with reranker computation costs, FVR-on/off comparisons, Gemini comparisons, and complete-evidence evaluation for Composite and Multi-hop questions.
\end{enumerate}

\section{Problem Setting and Research Questions}
Let $\mathcal{C}$ denote the corpus of source chunks in \dhara{} Bench and $\mathcal{G}_q\subseteq\mathcal{C}$ the annotated relevant chunks for a question $q$. The retrieval task is to return a ranked list of chunks from $\mathcal{C}$. For an EN record $x$, $P(x)\subseteq\mathcal{C}$ denotes its linked source chunks. The distinct source chunks collected before reranking form the \emph{candidate pool}; the final ranking is evaluated against $\mathcal{G}_q$.

Composite and Multi-hop questions also have annotated evidence groups, allowing evaluation of both individual relevant chunks and recovery of at least one complete supporting group.

The study addresses five research questions:
\begin{description}[style=nextline,leftmargin=0pt,labelwidth=0pt,labelsep=0pt]
\item[RQ1 - Direct retrieval and resource use.] How do SV and MV retrieval compare with the BM25 baseline across question categories and retrieval depths, and what are the neural models' latency and memory footprints for chunk storage?
\item[RQ2 - Value of Entity Notes.] What does adding EN contribute when the combined ranking retains the same number of candidates as direct retrieval?
\item[RQ3 - Reranker quality and time cost.] How do Qwen and BGE differ in retrieval quality and recorded computation time when applied to the same candidate pools?
\item[RQ4 - Contribution of verse retrieval.] How does enabling FVR change the final ranking, especially for Verse \& Textual Retrieval questions?
\item[RQ5 - Comparison with Gemini.] How do the combined EN and direct pipelines, using either reranker, compare with Gemini in category-wise and complete-evidence retrieval?
\end{description}

\section{Related Work}
\subsection{Lexical, single-vector, and multivector passage retrieval}
BM25 ranks passages through lexical matching~\cite{robertson2009bm25}. We use the established \dhara{} Bench baseline~\cite{dharaBenchmark2026} as the reference for the direct neural retrieval comparison.
Karpukhin et al.\ demonstrated dense passage retrieval (DPR) for open-domain question answering using separate question and passage encoders \cite{karpukhin2020dpr}. ColBERT and ColBERTv2 use token-level late interaction to retain fine-grained matching signals while supporting efficient passage search \cite{khattab2020colbert,santhanam2022colbertv2}. BGE-M3 supports dense, sparse, and multivector retrieval across languages and input lengths \cite{chen2024m3}. Thakur et al.\ demonstrate variation in retrieval effectiveness across datasets and domains through BEIR \cite{thakur2021beir}, highlighting the need for evaluation on the target collection.

\subsection{Neural passage reranking}
Neural reranking scores passages returned by a first-stage retriever through direct question-passage interaction. BERT-based passage reranking demonstrated the effectiveness of this approach \cite{nogueira2019bert}. BGE-Reranker-v2-M3 is a multilingual relevance-scoring model \cite{bgeRerankerV2M3}, distinct from the BGE-M3 embedding model used for first-stage retrieval. Qwen3 reranking models score the relevance of candidate passages to a query \cite{zhang2025qwen3}.

\subsection{Generated summaries and entity-based retrieval}
Generated representations provide alternative units for retrieving information from a corpus. Dense X Retrieval examines retrieval granularity and introduces self-contained propositions as retrieval units \cite{chen2024densex}. RAPTOR recursively clusters and summarizes passages to form a tree of representations at multiple levels of abstraction \cite{sarthi2024raptor}. GraphRAG builds an entity graph and generates community summaries to support questions requiring information from across a corpus \cite{edge2024graphrag}.

Entity-centred methods are particularly relevant to Entity Notes. EXPLAIN extracts and links entities, generates entity summaries, and retrieves those summaries alongside document chunks for answer generation \cite{liang2025explain}. EnSI-RAG constructs structured entity records linked to original passages and uses these records to locate evidence for long-document question answering \cite{meng2026ensirag}. These approaches establish entity-based records as an additional route to source passages.

Within this line of work, \dhara{} organizes entity descriptions as CEN and BEN, with links to source chunks for reranking. Our study examines this design within an IKS pipeline that also includes direct retrieval and FVR. Table~\ref{tab:insight-positioning} compares EN with the retrieval representations and organization of the related approaches.

\begin{table*}[!htbp]
\caption{Architectural comparison of generated retrieval representations and Entity Notes.}
\label{tab:insight-positioning}
\centering
\small
\begin{tabular}{@{}p{0.17\textwidth}p{0.25\textwidth}p{0.23\textwidth}p{0.27\textwidth}@{}}
\toprule
Approach & Derived retrieval representation & Organization / scope & Use of retrieved material \\
\midrule
Dense X Retrieval \cite{chen2024densex} & Self-contained propositions & Fine-grained retrieval units & Retrieved propositions supply information for downstream tasks \\
RAPTOR \cite{sarthi2024raptor} & Recursive abstractive summaries & Tree spanning multiple abstraction levels & Chunks and summaries provide context for answer generation \\
GraphRAG \cite{edge2024graphrag} & Community summaries & Communities in an entity graph & Community summaries support partial responses combined into an answer \\
EXPLAIN \cite{liang2025explain} & Entity summaries & Summaries of linked entities & Summaries and document chunks provide retrieved answer context \\
EnSI-RAG \cite{meng2026ensirag} & Structured entity records & Query-independent entity index & Records identify original passages used for answer synthesis \\
\dhara{} & Entity Notes (EN) & CEN + BEN & Notes identify source chunks for reranking and evaluation against \dhara{} Bench annotations \\
\bottomrule
\end{tabular}
\end{table*}

\section{The \dhara{} Evidence Engine}\label{sec:engine}
\subsection{Retrieval pipeline and source-chunk flow}
\begin{figure*}[!htbp]
\centering
\begin{tikzpicture}[
 font=\small,box/.style={draw,rounded corners=2pt,align=center,text width=43mm,minimum height=12mm,inner sep=4pt,outer sep=0pt},
 arrow/.style={-{Latex[length=2.2mm,width=1.6mm,sep=0pt]},line width=.8pt,shorten <=0pt,shorten >=0pt}]
\node[box] (q) at (0,0) {Question};
\node[box] (en) at (-4,-1.9) {Top 20 CEN/BEN records\\Map to $S_q$ source chunks};
\node[box] (sv) at (1,-1.9) {Direct retrieval\\$S_q$ source chunks};
\node[box] (fvr) at (6,-1.9) {Extract verse fragments\\FVR and source mapping};
\node[box] (er) at (-4,-3.6) {Rerank EN-linked chunks\\BGE or Qwen};
\node[box] (sr) at (1,-3.6) {Rerank direct chunks\\Same reranker};
\node[box,text width=61mm] (merge) at (-1.5,-5.3) {CombMAX on existing scores\\Retain top $S_q$ chunks};
\node[box,text width=61mm] (final) at (-1.5,-7.1) {Prepend FVR source chunks\\Deduplicate; retain final top 20};
\coordinate (fork) at ($(q.south)+(0,-.35)$);
\draw[line width=.8pt] (q.south) -- (fork);
\draw[line width=.8pt] (en.north |- fork) -- (fvr.north |- fork);
\draw[arrow] (en.north |- fork) -- (en.north);
\draw[arrow] (sv.north |- fork) -- (sv.north);
\draw[arrow] (fvr.north |- fork) -- (fvr.north);
\draw[arrow,dashed] (en.east) -- node[above,font=\scriptsize] {$S_q$} (sv.west);
\draw[arrow] (en.south) -- (er.north);
\draw[arrow] (sv.south) -- (sr.north);
\draw[arrow] (er.south) -- (er.south |- merge.north);
\draw[arrow] (sr.south) -- (sr.south |- merge.north);
\draw[arrow] (merge.south) -- (final.north);
\draw[arrow] (fvr.south) |- (final.east);
\end{tikzpicture}
\Description{EN and direct retrieval each supply S q chunks. CombMAX combines their chunks using existing reranker scores. A source chunk retrieved by both branches appears once, with the higher of its two scores. The combined ranking is truncated to S q chunks. Optional FVR matches are prepended before the final twenty-result cutoff.}
\caption{Component pipeline. Both branches and the combined ranking retain $S_q$ candidates before the final cutoff and optional FVR.}
\label{fig:architecture}
\end{figure*}
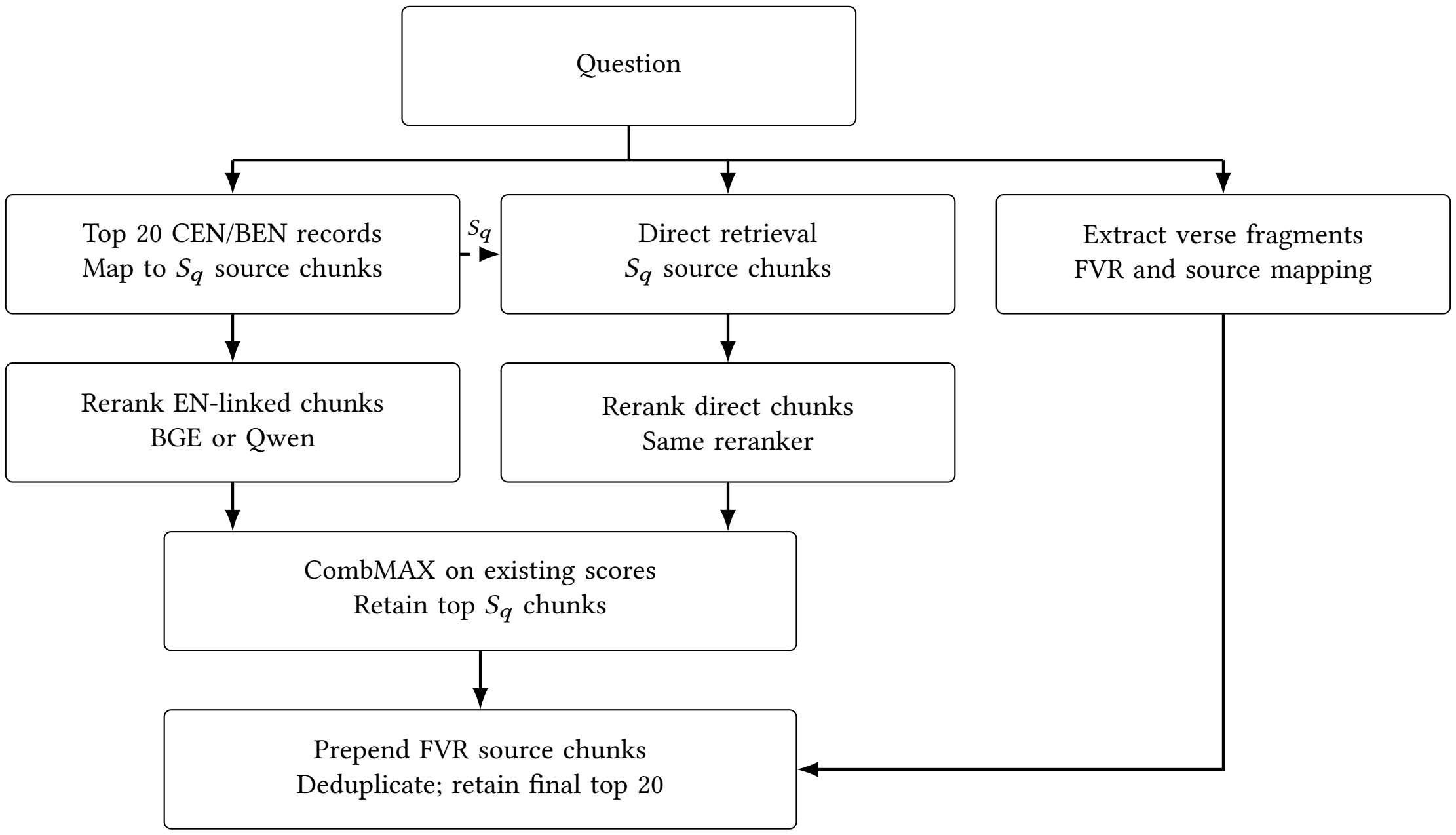

Figure~\ref{fig:architecture} summarizes the component pipeline used in Experiment 2. Experiment 1 compares direct SV and MV retrieval with a standalone BM25 baseline. In the component pipeline, EN retrieval identifies a pool of $S_q$ distinct provenance-linked source chunks. Direct retrieval supplies a second pool with the same number of candidates. Each pool is reranked separately using the same reranker. Source chunks from both branches are combined using CombMAX~\cite{fox1994combination}, which ranks distinct chunks using their highest existing reranker score. When the same source chunk is retrieved by both branches, it appears once in the combined ranking, with the higher of its two reranker scores. We retain the top $S_q$ chunks so that the combined ranking matches the direct branch's candidate count. FVR identifies source chunks associated with verse fragments and gives them priority in the final ranking. The output contains at most 20 distinct source chunks. Table~\ref{tab:storeinventory} reports the retrieval-store inventory.

\subsection{Neural embedding models}
We use Weaviate as the vector database to store and index source chunks and their embeddings, supporting hybrid retrieval through vector similarity and lexical matching~\cite{weaviateDocumentation}.

We consider two single-vector encoders, BGE-M3 and all-MiniLM-L6-v2, and the multivector model ColBERTv2. Single-vector encoders represent each text with one dense vector, while ColBERTv2 retains token-level vectors for fine-grained matching between questions and passages.

\subsubsection{BGE-M3}
BGE-M3 supports dense, sparse, and multivector representations across languages \cite{chen2024m3}. We encode each query and searchable record as one 1,024-dimensional vector using its dense output.

\subsubsection{all-MiniLM-L6-v2}
The all-MiniLM-L6-v2 encoder \cite{allMiniLML6v2} maps queries and passages to normalized 384-dimensional dense vectors. It provides a compact single-vector representation whose semantic similarity can be measured using cosine similarity.

\subsubsection{ColBERTv2}
ColBERTv2 uses contextualized token representations and late interaction to match queries with passages \cite{khattab2020colbert,santhanam2022colbertv2}. Each text is represented by a sequence of 128-dimensional token vectors. Its MaxSim formulation matches each query token to its most similar passage token and sums these similarities:
\begin{equation}
  s_{\mathrm{MaxSim}}(Q,D)=\sum_i\max_j \operatorname{sim}(q_i,d_j).
\end{equation}
Here, $Q=(q_i)$ and $D=(d_j)$ are the query and passage token-vector sequences, and $\operatorname{sim}$ denotes token-vector similarity.

\subsection{Entity Notes (EN): construction and source provenance}
Our hypothesis is that Entity Notes (EN) can improve retrieval by providing entity-focused descriptions that help locate relevant source chunks. Chapter-level notes describe an entity within a source unit, while book-level notes bring together its descriptions across chapters. These notes provide an additional way to match a question to relevant evidence. Each note retains links to its underlying source chunks, which are used for reranking and evaluation. We assess EN's contribution by combining these linked chunks with those returned by direct retrieval, retaining the same number of combined candidates as direct retrieval.

EN records are generated during corpus construction. Chapter-level Entity Notes (CEN) are consolidated into Book-level Entity Notes (BEN) by grouping records with the same exact entity name across chapters of one book. The entity-extraction, CEN-generation, and BEN-consolidation prompts are given in Supplementary Material, Sections~\ref*{supp-app:entity-extraction}, \ref*{supp-app:ces-generation}, and~\ref*{supp-app:bes-consolidation}, respectively; generation settings are consolidated in Supplementary Material, Section~\ref*{supp-app:generation-settings}.

\subsubsection{Entity extraction and chapter-level notes}
For each chapter or source unit, a locally hosted Qwen3-14B model identifies named entities, preserving their forms as they appear in the text \cite{yang2025qwen3}. These include persons, places, organizations, trees, rivers, and festivals.

Qwen3-14B then receives the chapter text, its source reference, and batches of extracted entity names. The CEN prompt requests a description grounded in the chapter, answering ``What?'' and ``How?'' in approximately four or five sentences and beginning with the entity and source reference. We retain the chapter content and reference with the generated response so that each CEN can be mapped back to the source material from which it was produced.

Let $C_{b,h,e}$ denote a CEN for entity $e$ derived from chapter $h$ of book $b$. Its provenance mapping is
\begin{equation}
  P(C_{b,h,e}) \subseteq \mathcal{C}.
\end{equation}
Here, $\mathcal{C}$ is the full corpus of source chunks, and $P(C_{b,h,e})$ is the subset linked to the CEN record for entity $e$ in chapter $h$ of book $b$.

\subsubsection{Book-level consolidation and source mapping}
We group CEN records across chapters of the same book by \emph{exact entity-name match}. We do not merge aliases or spelling variants at this stage. When several CEN records share the same entity string, Qwen3-14B receives their descriptions and references and produces one cohesive Book-level Entity Note (BEN). The prompt asks the model to retain the supplied information, use numbered inline citations, and end with the corresponding numbered reference list.

Let $B_{b,e}$ denote the BEN for entity $e$ in book $b$, and let $\mathcal{A}(B_{b,e})$ be the set of CEN records consolidated into that BEN. Its linked source chunks are given by
\begin{equation}
  P(B_{b,e})=\bigcup_{C \in \mathcal{A}(B_{b,e})}P(C).
\end{equation}
Here, $C$ denotes a constituent CEN record and $P(C)$ its linked source chunks. The BEN inherits all these links, with each source chunk included once.

The resulting provenance path is
\begin{equation}
  \text{BEN} \rightarrow \text{CEN} \rightarrow \text{Source Chunks}.
\end{equation}
At query time, the EN branch retrieves the top 20 CEN/BEN records and maps them to distinct source chunks. Every EN record in Table~\ref{tab:storeinventory} links to at least one chunk. Since a record can link to several chunks, the resulting candidate pool can exceed 20.

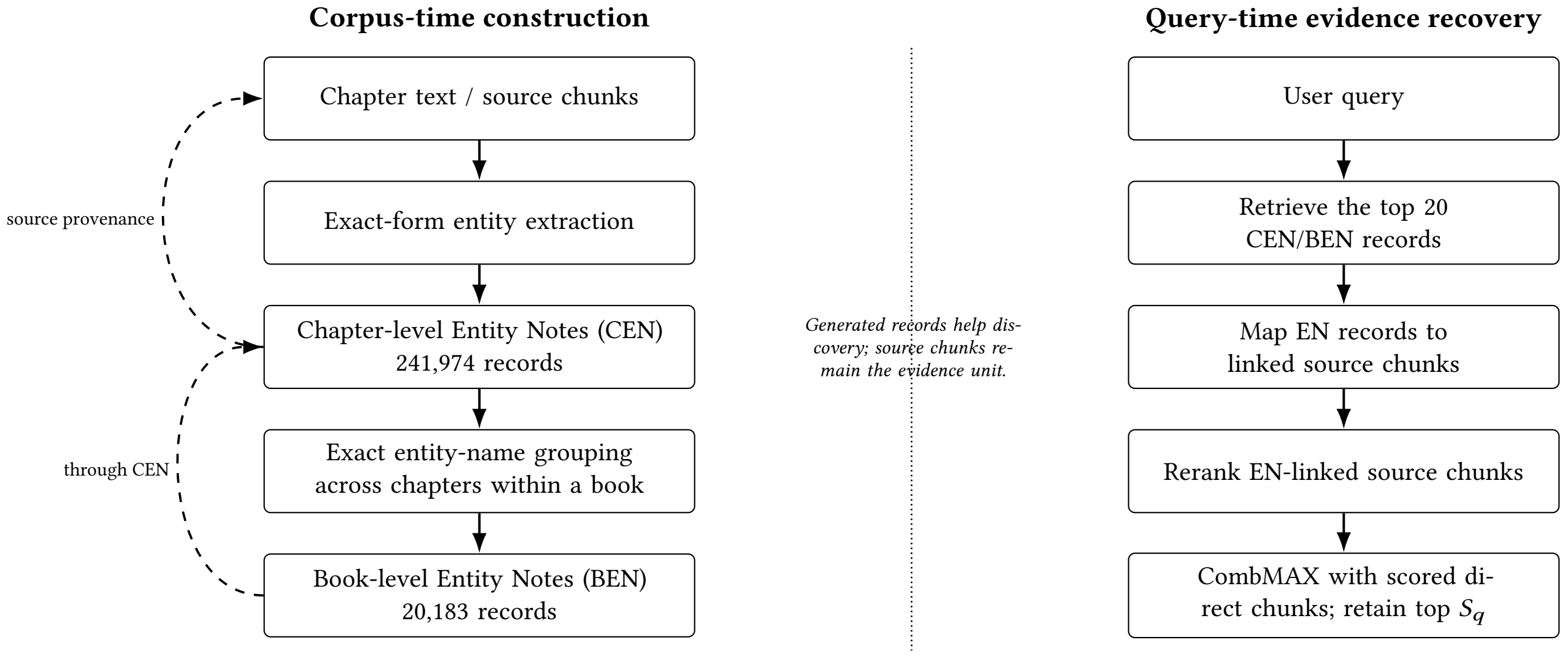
\begin{figure*}[!htbp]
\centering
\resizebox{0.94\textwidth}{!}{%
\begin{tikzpicture}[
  font=\small,
  box/.style={draw, rounded corners=2pt, align=center, minimum height=9mm, text width=44mm, inner sep=4pt},
  arrow/.style={-{Latex[length=2.2mm,width=1.6mm]}, line width=0.75pt},
  prov/.style={-{Latex[length=2.2mm,width=1.6mm]}, dashed, line width=0.7pt},
  phase/.style={font=\bfseries}
]
\node[phase] at (-4.7,0.85) {Corpus-time construction};
\node[phase] at (4.7,0.85) {Query-time evidence recovery};
\draw[densely dotted, line width=0.5pt] (0,0.55) -- (0,-6.0);

% Corpus-time: generation creates retrieval aids while preserving backward links.
\node[box] (source)  at (-4.7, 0.0) {Chapter text / source chunks};
\node[box] (entity)  at (-4.7,-1.35) {Exact-form entity extraction};
\node[box] (facts)   at (-4.7,-2.70) {Chapter-level Entity Notes (CEN)\\241,974 records};
\node[box] (group)   at (-4.7,-4.05) {Exact entity-name grouping across chapters within a book};
\node[box] (summary) at (-4.7,-5.40) {Book-level Entity Notes (BEN)\\20,183 records};
\draw[arrow] (source.south) -- (entity.north);
\draw[arrow] (entity.south) -- (facts.north);
\draw[arrow] (facts.south) -- (group.north);
\draw[arrow] (group.south) -- (summary.north);
\draw[prov] (facts.west) to[out=180,in=180,looseness=1.35] node[left, font=\scriptsize] {source provenance} (source.west);
\draw[prov] (summary.west) to[out=180,in=180,looseness=1.15] node[left, font=\scriptsize] {through CEN} (facts.west);

% Query-time: retrieve derived records, then return to canonical source evidence.
\node[box] (query)  at (4.7, 0.0) {User query};
\node[box] (iret)   at (4.7,-1.35) {Retrieve the top 20 CEN/BEN records};
\node[box] (map)    at (4.7,-2.70) {Map EN records to linked source chunks};
\node[box] (union)  at (4.7,-4.05) {Rerank EN-linked source chunks};
\node[box] (rerank) at (4.7,-5.40) {CombMAX with scored direct chunks; retain top $S_q$};
\draw[arrow] (query.south) -- (iret.north);
\draw[arrow] (iret.south) -- (map.north);
\draw[arrow] (map.south) -- (union.north);
\draw[arrow] (union.south) -- (rerank.north);

\node[align=center, text width=31mm, font=\scriptsize\itshape] at (0,-2.70)
{Generated records help discovery; source chunks remain the evidence unit.};
\end{tikzpicture}%
}
\Description{Two-phase diagram of Entity Notes. The left panel shows corpus-time construction from source chunks through exact-form entity extraction, Chapter-level Entity Notes (CEN), exact entity-name grouping, and Book-level Entity Notes (BEN), with dashed provenance links back toward source material. The right panel shows query-time retrieval of CEN or BEN records, provenance projection to source chunk identifiers, reranking, and merging with scored direct source chunks.}
\caption{Entity Note construction and retrieval. CEN records are consolidated into BEN records; retrieved notes identify linked source chunks for reranking.}
\label{fig:insights-lifecycle}
\end{figure*}

Figure~\ref{fig:insights-lifecycle} separates corpus-time construction from query-time retrieval and source mapping.

\begin{table}[!htbp]
\caption{Retrieval-store inventory.}
\label{tab:storeinventory}
\centering
\small
\begin{tabular}{@{}lr@{}}
\toprule
Record type & Count \\
\midrule
Source chunks & 60,121 \\
Chapter-level Entity Notes (CEN) & 241,974 \\
Book-level Entity Notes (BEN) & 20,183 \\
Parent verses & 165,101 \\
Generated verselets & 715,158 \\
Verse source records & 15 \\
\bottomrule
\end{tabular}
\end{table}

\subsection{Fuzzy Verse Retrieval (FVR)}
FVR retrieves verses from partial fragments through script normalization, spelling-variant generation, substring matching, and trigram similarity. Matched verses are associated with source chunks for inclusion in the final ranking.

We construct a dedicated verse database from the verse-bearing sources represented in the \dhara{} Bench corpus~\cite{dharaBenchmark2026}. The source works are \href{https://sanskritsahitya.org/mahabharatam}{Mah\={a}bh\={a}rata}, \href{https://valmiki.gitasupersite.in/}{V\={a}lm\={i}ki R\={a}m\={a}ya\d{n}a}, \href{https://github.com/sanskrit/raw_etexts/blob/master/purANam/mAdhva-app/agni-purANam.md}{Agni Pur\={a}\d{n}a}, \href{https://github.com/sanskrit/raw_etexts/blob/master/kalpaH/ApastambaH/apastamba_dharma_sutra.md}{\={A}pastamba Dharma-s\={u}tra}, \href{https://github.com/sanskrit/raw_etexts/tree/master/AyurvedaH/charakaH}{Caraka Sa\d{m}hit\={a}}, \href{https://www.wisdomlib.org/hinduism/book/kathasaritsagara-sanskrit}{Kath\={a}sarits\={a}gara}, \href{https://sanskritdocuments.org/sanskrit/natyashastra/}{N\={a}\d{t}ya\'{s}\={a}stra}, \href{https://github.com/sanskrit/raw_etexts/tree/master/AyurvedaH/aShTAnga-hRdayam}{A\d{s}\d{t}\={a}\.{n}ga H\d{r}dayam}, \href{https://old.gitasupersite.in/srimad?language=dv&field_chapter_value=1&field_nsutra_value=1&setgb=1&etradi=1&choose=1}{Bhagavadg\={i}t\={a}}, \href{https://github.com/sanskrit/raw_etexts/blob/master/vedAntam/advaitam/advaita-shAradA/bhAShya/Kathaka.md}{Ka\d{t}ha Upani\d{s}ad}, \href{https://www.ebharatisampat.in/readbook3.php?bookid=ODQ1MzgzNzg1NjQ4NDQ0&pageno=MjI0MjQyNjk5NTk=}{M\={a}nas\={a}ra}, \href{https://github.com/sanskrit/raw_etexts/tree/master/vedaH/Rg/shakala/saMhitA/sAyaNabhAShyam}{\d{R}g \'{S}\={a}kala Sa\d{m}hit\={a}}, and \href{https://old.gitasupersite.in/yogasutra_content?language=dv&field_chapter_value=1&field_nsutra_value=1&enable_sutra=1&enable_bhaysa=1&enable_vritti=1}{Yoga-s\={u}tra}.

The database contains 165,101 complete verses stored under 15 source records (Table~\ref{tab:storeinventory}). These verses form the basis for verselet construction, described below. Complete verses and verselets are stored in PostgreSQL 15.5 together with their source metadata and links between each verselet and its parent verse.

\subsubsection{Detecting verse fragments in questions}
\dhara{} uses Qwen3-4B-Instruct-2507 \cite{qwen3Instruct2507} to extract quoted or partially remembered text from verses, shlokas, mantras, compositions, and songs in a question. The extractor distinguishes these textual fragments from topic names and requests for definitions, and can identify multiple fragments in the same question. FVR is activated when at least one fragment is detected. Supplementary Material, Section~\ref*{supp-app:verse-extraction} provides the extraction prompt, and Supplementary Table~\ref*{supp-tab:generationsettings} lists the model settings. The FVR-on/off comparisons in Section~\ref{sec:verseablationdesign} assess the contribution of this component to retrieval effectiveness.

\subsubsection{Verselet length selection and construction}
A \emph{verselet} is a short, searchable segment of a verse. We divide verses into these smaller units to retrieve complete verses from partial fragments.

\begin{figure*}[!htbp]
\centering
\includegraphics[width=\linewidth]{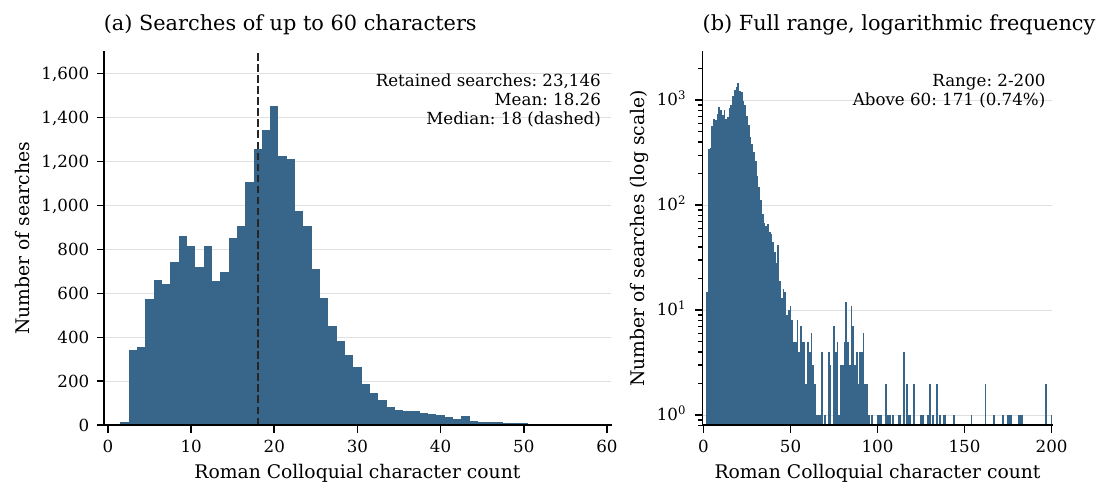}
\Description{Two histograms show verse-search lengths after conversion to Roman Colloquial. The left panel shows lengths up to 60 characters, with a peak at 20 and a median of 18. The right panel shows the full range of 2 to 200 characters on a logarithmic frequency scale.}
\caption{Verse-search lengths in Roman Colloquial (23,146 searches; one-character bins). Left: lengths through 60 characters. Right: the full 2-200-character range, with logarithmic frequency.}
\label{fig:versequerylength}
\end{figure*}

To guide the segment length, we examined 23,146 retained verse searches from a population of 676 registered users of the \dhara{} application\footnote{\dhara{} application: \url{https://dhara.bheri.in/}.}. We first converted searches in different scripts to Roman Colloquial using Aksharamukha \cite{rajanAksharamukha}. Their mean length was 18.26 characters, the median was 18, and the most frequent length was 20 (Figure~\ref{fig:versequerylength}); lengths ranged from 2 to 200.

A Devanagari \emph{akshara} is a written syllabic unit: an independent vowel or a consonant or consonant cluster carrying an inherent or explicitly marked vowel, together with associated signs~\cite{unicode2024devanagari}. One akshara can comprise multiple Unicode characters. At approximately 2.4-2.5 Roman Colloquial characters per Devanagari akshara, the most frequent length corresponds to about eight aksharas. This provided a practical guide for choosing short verselets with a target length of ten units and an accepted range of 8-12.

Construction begins by cleaning the verse text and splitting it at danda punctuation marks. Cleaning removes Latin letters, configured Vedic marks, punctuation, and numerals, and normalizes spacing. To estimate akshara length, we transliterate the Devanagari text to IAST and count vowel units, treating \texttt{ai} and \texttt{au} as one unit each. Segments of 8-12 units become individual verselets. Longer segments are divided near the ten-unit target, splitting words when necessary, and short remainders are joined to adjacent segments. Construction produces 715,158 verselets (Table~\ref{tab:storeinventory}), with at least one verselet for every parent verse.

\subsubsection{Fragment normalization and spelling variants}
Indic terms can have several spellings when written in English; for example, Srinivas may be written as Shrinivas, Sreenivas, or Shreenivas. FVR generates spelling variants to accommodate such differences when matching a verse fragment to the stored text.

Before searching, FVR cleans each extracted fragment, converts it to Roman Colloquial, lowercases it, and removes spaces. A replacement table generates alternative strings by optionally applying spelling substitutions and their combinations. For example, \texttt{agnimeele purohitam} becomes \texttt{agnimeelepurohitam}; the rule \texttt{ee}$\rightarrow$\texttt{i} also produces \texttt{agnimilepurohitam}. The normalized original is searched first, followed by the generated variants. These are spelling-based search candidates, not synonyms. Supplementary Material, Section~\ref*{supp-app:verse-implementation} summarizes the replacement rules.

\subsubsection{Substring and trigram matching}
FVR finds candidate verses in two ways. \emph{Substring matching} finds verselets that contain the normalized fragment or one of its spelling variants. \emph{Trigram matching} compares the original normalized fragment with stored verselets using overlapping sequences of three characters, allowing similar fragments to match even when their spelling differs. Trigram candidates are retained when their PostgreSQL similarity score exceeds 0.2.

Substring matches take priority. They are processed in spelling-variant order and then by verselet position within the verse. Trigram matches follow in decreasing similarity order. Matches that belong to the same parent verse are combined, and each fragment lookup returns at most 15 distinct parent verses. The matching scores determine retrieval order; they are not probabilities of relevance.

For a fragment $f$, this procedure is written as
\begin{equation}
R_{\mathrm{FVR}}(f)=\operatorname{Top}_{15}\!\left(\operatorname{DedupVerse}\left(R_{\mathrm{sub}}(f)\ \Vert\ R_{\mathrm{tri}}(f)\right)\right),
\end{equation}
where $R_{\mathrm{sub}}$ contains the ordered substring matches and $R_{\mathrm{tri}}$ contains the trigram matches above the threshold. The operator $\Vert$ places substring matches before trigram matches, and $\operatorname{DedupVerse}$ retains each parent verse once.

\subsubsection{Mapping matched verses to source chunks}\label{sec:versechunkmapping}
During corpus preparation, translation and commentary chunks are associated with their corresponding verses, and the verse text is stored in the chunk metadata. Each verselet retains links to its complete parent verse and source record:
\begin{equation}
\text{Verselet}\rightarrow\text{Parent verse}\rightarrow\text{Source record}.
\end{equation}
To link a verse to its source chunks, we compare the complete verse with the verse text stored in each chunk's metadata. RapidFuzz measures how closely the texts match. We keep chunks with a similarity score of at least 85 out of 100 and place the closest matches first.

Verse-to-chunk mappings are precomputed for all 165,101 verses and stored with each verse in PostgreSQL as ranked chunk IDs. At query time, FVR retrieves the matched verse's stored mapping and fetches the associated source chunks. A verse with an empty mapping contributes no source chunks. Supplementary Material, Section~\ref*{supp-app:verse-implementation} provides the normalization details and mapping coverage.

\subsection{Reranking and final ranking}
\subsubsection{Matching candidate counts between retrieval branches}
The top 20 CEN/BEN records identify a set $E_q$ of distinct source chunks. We define $S_q=|E_q|$ and retrieve $S_q$ source chunks directly, forming $D_q$. This gives both branches the same candidate count before reranking. After combining both branches with CombMAX, we retain the top $S_q$ chunks. The combined candidate count therefore matches that of direct retrieval.

\subsubsection{Reranking and CombMAX fusion}
Either BGE-Reranker-v2-M3~\cite{bgeRerankerV2M3} or Qwen3-Reranker-4B~\cite{zhang2025qwen3} scores both pools against the question in separate calls. BGE returns relevance scores directly. Qwen assigns each passage a relevance score based on how likely it is to answer ``yes'' when asked whether the passage is relevant to the question. Both rerankers are served through vLLM on the DGX Spark using BF16 precision. The same model is used for both branches within a configuration; scores from different rerankers are never mixed.

We combine the two reranked lists using CombMAX~\cite{fox1994combination}. If both lists contain the same source chunk, we keep it once with the higher of its two scores. We then sort the chunks from highest to lowest score and retain the top $S_q$, matching the number of candidates in direct retrieval. Each metric uses the first few chunks required by its cutoff, such as five for Precision@5 and twenty for Recall@20. Combining the lists reuses the existing scores and does not require reranking again. Supplementary Table~\ref*{supp-tab:auditsettings} gives the batching and preprocessing settings, and Supplementary Material, Section~\ref*{supp-app:qwen-reranking} provides the Qwen prompt.

\subsubsection{Adding FVR matches}
When FVR is enabled, verse-matched source chunks are placed first, followed by the CombMAX ranking. Duplicates are removed, and at most 20 distinct chunks are returned. If FVR finds no matching chunks, the ranking remains unchanged.

\section{Experimental Design}\label{sec:experimental}
\begin{figure*}[!htbp]
\centering
\begin{tikzpicture}[
 font=\small,
 box/.style={draw,rounded corners=2pt,align=center,text width=65mm,minimum height=13mm,inner sep=4pt},
 arrow/.style={-{Latex[length=2.2mm,width=1.6mm]},line width=0.8pt}
]
\node[box,text width=90mm] (inputs) at (0,0)
 {\dhara{} Bench evaluation set: 325 questions\\Associated corpus: 60,121 source chunks};
\node[box] (rep) at (-4.1,-2.6)
 {\textbf{Experiment 1: representations}\\Hybrid: BGE-M3 / all-MiniLM-L6-v2\\ColBERTv2; BM25 baseline\\Rankings through depth 1,000};
\node[box] (sv) at (4.1,-2.6)
 {\textbf{Experiment 2: components}\\Representation selected from Experiment 1\\Top 20 CEN/BEN records\\Map to $S_q$ chunks; match direct budget};
\coordinate (fork) at (0,-.65);
\draw[arrow] (inputs.south) -- (fork) -| (-4.1,-.9) -- (rep.north);
\draw[arrow] (fork) -| (4.1,-.9) -- (sv.north);
\node[box] (prefix) at (-4.1,-4.4)
 {Evaluate the top-ranked results\\at different retrieval depths};
\node[box] (components) at (4.1,-4.4)
 {Rerank branches with BGE or Qwen\\CombMAX; retain top $S_q$\\Optional FVR priority\\Remove duplicates; retain final top 20};
\draw[arrow] (rep.south) -- (prefix.north);
\draw[arrow] (sv.south) -- (components.north);
\node[box] (rep_eval) at (-4.1,-6.2)
 {Question-category-wise retrieval metrics\\Shared and unique relevant chunks\\Retrieval latency and memory footprint\\(chunk storage on disk)};
\node[box] (sv_eval) at (4.1,-6.2)
 {Question-category-wise metrics and timing\\EN addition; BGE/Qwen; FVR on/off\\Compare EN+Direct+FVR with Gemini};
\draw[arrow] (prefix.south) -- (rep_eval.north);
\draw[arrow] (components.south) -- (sv_eval.north);
\end{tikzpicture}
\Description{Two experiment paths share the Dhara Bench evaluation set and source corpus. Experiment 1 compares SV, MV, and BM25 retrieval using 1,000-result rankings and evaluates ranking prefixes, relevant-chunk overlap, and latency and memory footprint for chunk storage on disk. Experiment 2 examines components and controls using the representation selected from Experiment 1, with Gemini as a comparator.}
\caption{Experiment 1 compares direct retrieval with BM25 and informs the representation choice for Experiment 2. Four component comparisons examine EN, rerankers, FVR, and Gemini.}
\label{fig:experiment-setup}
\end{figure*}
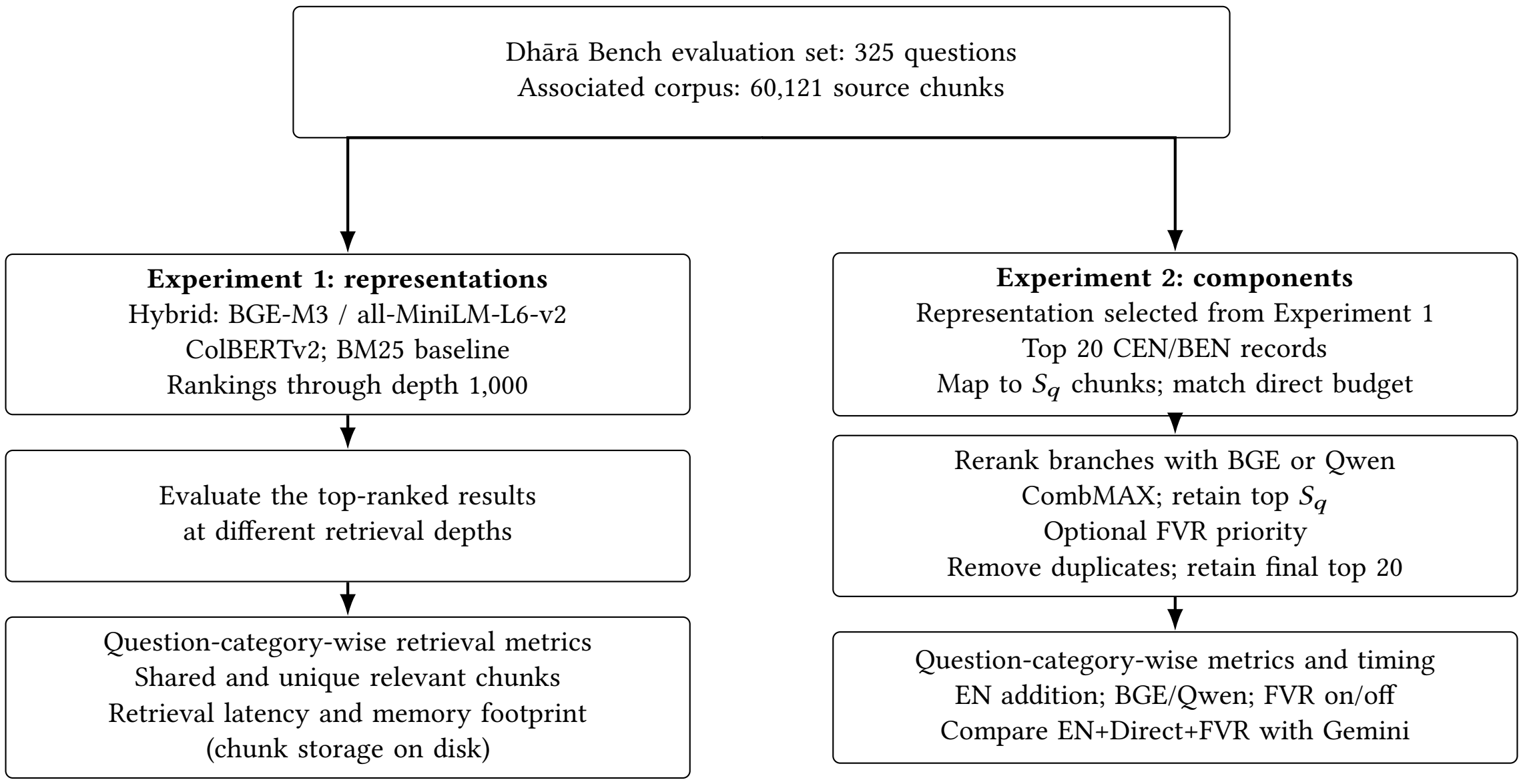

Figure~\ref{fig:experiment-setup} summarizes the representation comparison in Experiment 1 and the component comparison in Experiment 2. Experiment 1 informs the choice of retrieval representation for the subsequent component study.

\subsection{Evaluation set, source corpus, and question categories}
Both experiments use all \NQuestions{} questions in \dhara{} Bench and its associated corpus of 60,121 source chunks~\cite{dharaBenchmark2026}. The benchmark annotations identify the source chunks relevant to each question. The standalone BM25 baseline is taken unchanged from \dhara{} Bench~\cite{dharaBenchmark2026} and serves as the lexical reference in Experiment 1.

Table~\ref{tab:categories} lists the twelve question categories and their sizes. Following the \dhara{} Bench evaluation protocol~\cite{dharaBenchmark2026}, we average scores across all questions in each category, including zero scores for questions with no retrieved results.

\begin{table*}[!htbp]
\caption{Question categories in the 325-question \dhara{} Bench evaluation set.}
\label{tab:categories}
\centering\small
\begin{tabular}{@{}lp{0.70\textwidth}r@{}}
\toprule
Code & Question category & $n$ \\
\midrule
T01 & Fact Retrieval & 51 \\
T02 & List / Set Retrieval & 40 \\
T03 & Counting / Aggregation & 22 \\
T04 & Chronology \& Sequence understanding & 25 \\
T05 & Lexical Understanding & 31 \\
T06 & Composite & 19 \\
T07 & Multi-hop Reasoning & 15 \\
T08 & Verse \& Textual Retrieval & 26 \\
T09 & False Premise & 18 \\
T10 & Summarization / Expansion / Composition & 29 \\
T11 & Analytical / Interpretive Reasoning & 32 \\
T12 & Mathematical / Logical / Computational Problems & 17 \\
\bottomrule
\end{tabular}

\end{table*}

The 19 Composite (T06) and 15 Multi-hop (T07) questions also have annotations identifying alternative groups of source chunks that together provide complete evidence. These annotations contain 401 groups, including 35 single-chunk groups. Conventional retrieval metrics use the distinct relevant chunks across all groups; Section~\ref{sec:completedesign} defines complete-evidence scoring. The remaining categories use their annotated lists of relevant source chunks.

\subsection{Experiment 1: comparing retrieval representations}\label{sec:representationdesign}
\subsubsection{Models and retrieval depth}
Experiment 1 compares BGE-M3~\cite{chen2024m3} and all-MiniLM-L6-v2~\cite{allMiniLML6v2} single-vector retrieval, ColBERTv2~\cite{santhanam2022colbertv2} multivector retrieval, and the standalone BM25 baseline. Each neural model is used with hybrid search, combining vector similarity with lexical matching. The three neural configurations search collections containing the same 60,121 source chunks and are evaluated on the same questions and annotations in the \dhara{} Bench evaluation set. These neural-retrieval collections use cosine HNSW indexing~\cite{malkov2020hnsw}. Hybrid search uses chunk text, a vector weight of 0.7, and a lexical weight of 0.3. BGE-M3 and all-MiniLM-L6-v2 use normalized single vectors of 1,024 and 384 dimensions, respectively. ColBERTv2 uses 128-dimensional token vectors with cosine MaxSim aggregation.

Each neural configuration returns 1,000 ranked source chunks per question. We reuse the BM25 rankings from \dhara{} Bench, containing up to 1,000 source chunks per question. We evaluate retrieval at smaller cutoffs by taking the corresponding prefixes of each ranking. This experiment measures direct retrieval; the effects of Entity Notes, reranking, and FVR are examined separately in Experiment 2. Implementation settings are provided in Supplementary Material, Sections~\ref*{supp-app:direct-settings} and~\ref*{supp-app:minilm-settings}.

\subsubsection{Shared and unique relevant evidence}
We compare the top-20 BGE-M3 and ColBERTv2 rankings to examine whether they retrieve the same relevant source chunks. For each question, we count annotated relevant source chunks retrieved by both models, only by BGE-M3, and only by ColBERTv2. Counts are summed within each question category. A chunk relevant to multiple questions is counted separately for each question.

\subsubsection{BM25 lexical baseline from \dhara{} Bench}\label{sec:bm25design}
The standalone BM25 baseline is reused unchanged from \dhara{} Bench~\cite{dharaBenchmark2026,robertson2009bm25}. Its Pyserini/Anserini implementation searches source-chunk text with $k_1=0.9$, $b=0.4$, English stopword removal, and Porter stemming. We evaluate the saved rankings using the same questions, relevance annotations, and cutoffs as the neural configurations. Supplementary Material, Section~\ref*{supp-app:bm25-settings} provides the index and query settings. This standalone baseline is distinct from the lexical component of Weaviate hybrid search.

\subsection{Experiment 2: EN, reranking, FVR, and Gemini}\label{sec:budgetcontrol}
The representation selected from Experiment 1 supplies the direct retrieval branch. Four comparisons examine the contribution of EN, the quality and computation time of the rerankers, FVR's effect on the final ranking, and performance against Gemini. Table~\ref{tab:verifiedconfigs} defines the configurations. All final rankings contain at most 20 source chunks.

\begin{table*}[!htbp]
\caption{Component configurations. B/Q denotes BGE/Qwen reranking; $S_q$ is the EN-derived candidate count.}
\label{tab:verifiedconfigs}
\centering\small
\begin{tabularx}{\textwidth}{@{}lXc@{}}
\toprule
Configuration & Candidate selection and ranking & FVR \\
\midrule
Direct matched+B/Q & $S_q$ directly retrieved chunks, reranked & Off \\
EN+Direct+B/Q & Top $S_q$ chunks after CombMAX of the EN and direct rankings & Off \\
EN+Direct+B/Q+FVR & Same merged ranking, with FVR matches prepended & On \\
Gemini & Generated references mapped to chunks and manually reviewed & - \\
\bottomrule
\end{tabularx}
\end{table*}

\subsubsection{EN contribution at matched retained candidate counts}\label{sec:encontroldesign}
For each question, the top 20 CEN/BEN records are mapped through stored provenance links to $S_q$ distinct source chunks. Direct retrieval returns $S_q$ chunks for the matched branch. The candidate count has a mean of 671.35, a median of 369, and a range of 14-7,251. For two questions, $S_q$ is below 20 (14 and 17), so both matched-SV and EN+SV rankings contain 14 and 17 results, respectively, with FVR disabled.

The EN comparison uses CombMAX to combine the scored EN pool with this matched direct branch, using the same reranker and disabling FVR in both conditions. A source chunk retrieved by both branches is included once, with the higher of its two reranker scores. The combined chunks are sorted by score, and only the top $S_q$ are retained. Both EN+SV and direct retrieval therefore have the same retained candidate count. Metrics are computed from the leading results at the reported cutoffs. This comparison measures the value of EN in selecting the final candidates while holding their number fixed.

\subsubsection{Reranker quality and computation time}\label{sec:rerankerdesign}
We compare BGE and Qwen on the same EN and direct candidate sets, applying CombMAX and retaining the top $S_q$, and disabling FVR. Category-wise scores show the ranking differences, while recorded component times quantify the associated computation.

\subsubsection{FVR enabled and disabled}\label{sec:verseablationdesign}
For each reranker, FVR is applied to the same merged EN+Direct ranking used in the FVR-off condition. The pre-FVR list contains at most $\min(20,S_q)$ chunks and is identical within each pair. The comparison therefore measures the effect of prepending verse-associated chunks and removing duplicates at the final cutoff. We evaluate all 325 questions and emphasize Verse \& Textual Retrieval (T08).

\subsubsection{Comparison with Gemini}
The principal Gemini comparison uses both EN+Direct+B+FVR and EN+Direct+Q+FVR. The FVR-off comparisons are included in Supplementary Material, Section S2. All configurations are scored against the same source-chunk annotations; the Gemini mapping and review procedure is described next.

\subsection{Gemini as an LLM-based evidence retriever: source mapping and review}\label{sec:geminidesign}
We evaluate Gemini as a large language model (LLM)-based evidence retriever, prompting it to return ranked candidate passages and source references for each question. This comparison is motivated by the use of conversational models for information seeking: a large-scale study of ChatGPT identifies seeking information among its most common activities~\cite{chatterji2025chatgptusage}.

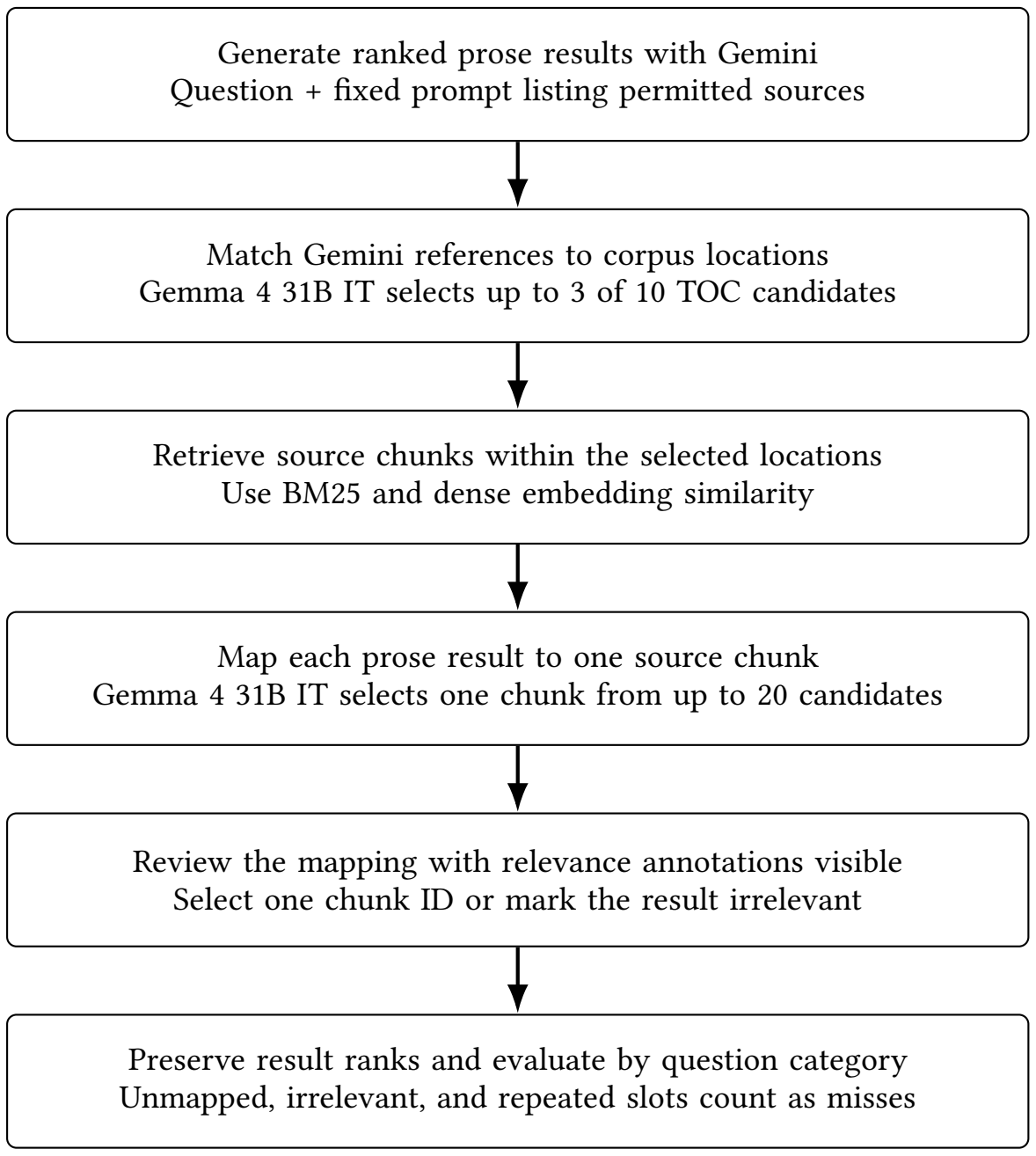
\begin{figure}[!htbp]
\centering
\begin{tikzpicture}[
 font=\footnotesize,
 box/.style={draw,rounded corners=2pt,align=center,text width=74mm,minimum height=10mm,inner sep=4pt},
 arrow/.style={-{Latex[length=2.2mm,width=1.6mm]},line width=0.8pt}
]
\node[box] (generation) at (0,0)
 {Generate ranked prose results with Gemini\\Question + fixed prompt listing permitted sources};
\node[box,below=5mm of generation] (locations)
 {Match Gemini references to corpus locations\\Gemma 4 31B IT selects up to 3 of 10 TOC candidates};
\node[box,below=5mm of locations] (retrieval)
 {Retrieve source chunks within the selected locations\\Use BM25 and dense embedding similarity};
\node[box,below=5mm of retrieval] (selection)
 {Map each prose result to one source chunk\\Gemma 4 31B IT selects one chunk from up to 20 candidates};
\node[box,below=5mm of selection] (review)
 {Review the mapping with relevance annotations visible\\Select one chunk ID or mark the result irrelevant};
\node[box,below=5mm of review] (evaluation)
 {Preserve result ranks and evaluate by question category\\Unmapped, irrelevant, and repeated slots count as misses};
\draw[arrow] (generation.south) -- (locations.north);
\draw[arrow] (locations.south) -- (retrieval.north);
\draw[arrow] (retrieval.south) -- (selection.north);
\draw[arrow] (selection.south) -- (review.north);
\draw[arrow] (review.south) -- (evaluation.north);
\end{tikzpicture}
\Description{A six-step flow shows Gemini prose generation, reference-based candidate-location identification and selection with Gemma 4 31B IT, source-chunk retrieval, single-chunk selection with the same model, manual review with evaluation annotations visible, and rank-preserving category-wise scoring.}
\caption{Gemini as an LLM-based evidence retriever: source mapping with Gemma 4 31B IT, manual review with relevance annotations visible, and scoring at preserved result ranks.}
\label{fig:gemini-evaluation-flow}
\end{figure}

Gemini generates prose passages and references, which we map to source chunks and manually review before evaluation. Figure~\ref{fig:gemini-evaluation-flow} summarizes this process: generate the results, locate the corresponding source material, select a source chunk, review the mapping, and score the resulting ranking.

\subsubsection{Generating prose results}
The \texttt{gemini-3.7-flash} model~\cite{gemini37Flash} receives a fixed prompt listing permitted sources and requesting ranked results in JSON. The prompt contains 20 prose-source entries and 13 verse-source entries. This comparison evaluates the ranked prose results, with up to 20 results per question, after mapping them to source chunks. Generation uses a cached static prompt and thinking level \texttt{HIGH}. Supplementary Material, Section~\ref*{supp-app:gemini-prompt} summarizes the prompt and source inventory, and Supplementary Material, Section~\ref*{supp-app:gemini-settings} records the generation settings. Gemini results are evaluated on the same \NQuestions{} questions in the \dhara{} Bench evaluation set.

\subsubsection{Mapping prose results to source chunks}
We map Gemini's generated passages and references to source chunks using a locally hosted Gemma 4 31B IT model (\path{google/gemma-4-31B-it})~\cite{gemma4ModelCard}, served through vLLM. The same model performs two selection stages at temperature 0.0.

First, we match Gemini's returned reference against references in the corpus table of contents (TOC) to identify up to ten candidate locations. Gemma selects up to three locations by comparing the work, edition, chapter, and other reference details, with the question and generated passage providing context.

Second, we retrieve candidate source chunks from the selected locations using BM25 and semantic similarity computed with all-MiniLM-L6-v2 embeddings~\cite{allMiniLML6v2}. Gemma then examines up to twenty candidate chunks and selects the one that best matches Gemini's generated passage, taking the question into account. Matching allows paraphrased passages and differences in reference wording. Supplementary Material, Sections~\ref*{supp-app:gemini-mapping} and~\ref*{supp-app:gemma-prompts} provide the generation settings and prompt summaries.

\subsubsection{Manual review and preservation of result ranks}
We manually reviewed Gemini's source mappings through an interface displaying the question, generated passage and reference, Gemma's automatic mapping, candidate source chunks, and the annotated relevant chunks from \dhara{} Bench. Each recorded review decision identified one relevant source chunk or marked the result irrelevant. Corrections could select a chunk outside the automatic candidate list. Manual selections replaced the automatic mappings; where no manual decision was recorded, Gemma's mapping was retained. Because relevance annotations were visible during review, the reported Gemini scores reflect generation followed by automatic mapping and manual review.

Scoring preserved each result's original rank. Unmapped and irrelevant results counted as misses, and repeated chunks received credit only at their first occurrence. Later duplicates counted as misses without shifting subsequent ranks. The evaluated rankings contain 2,528 positions: 1,951 mapped chunk positions and 577 misses. All 325 questions remain in the evaluation, including eleven with empty rankings that receive zero scores. Each ranking contains at most 20 positions.

\subsection{Retrieval metrics and category-wise evaluation}\label{sec:completedesign}
We adopt the retrieval and complete-evidence metrics described in \dhara{} Bench~\cite{dharaBenchmark2026} and follow its per-question scoring and category-averaging protocol.

\subsubsection{Retrieval metrics and cutoffs}
The main comparisons report nDCG@10, Precision@5, Recall@20, and MRR@20; F1@10 is reported in the supplementary results. These metrics describe different aspects of retrieval. Precision@5 measures the proportion of relevant source chunks among the first five results. Recall@20 measures the proportion of annotated relevant source chunks found within the first twenty. nDCG@10 gives greater credit to relevant chunks placed earlier in the ranking \cite{jarvelin2002cumulated}. MRR@20 measures how early the first relevant chunk appears, using its reciprocal rank and assigning zero when none appears within twenty results.

Scores are calculated for each question before averaging. F1@10 uses that question's Precision@10 and Recall@10, and precision divides by the requested cutoff even when fewer results are returned. nDCG uses binary relevance. In Experiment 1, we report overall scores for all five metrics at retrieval cutoffs of 3, 5, 8, 10, 20, 50, 100, 500, and 1,000. Additional category-wise plots in Supplementary Material, Section S1 show these metrics at cutoffs of 20, 50, 100, 500, and 1,000. At each cutoff, reciprocal rank receives credit only if the first relevant chunk appears within that cutoff.

\subsubsection{Comparing results within each category}
All methods are compared on the same questions and annotations in the \dhara{} Bench evaluation set. We calculate each category score by averaging the per-question scores in that category. To compare two configurations, we calculate their score difference for each question and average these differences within the category. A positive difference favors the first configuration in the comparison.

\subsubsection{Retrieving complete evidence groups}
For Composite (T06) and Multi-hop (T07) questions, complete-evidence scoring checks whether a ranking retrieves every chunk in at least one annotated evidence group. A group is complete when its last required chunk appears; when several alternative groups exist, we use the earliest completed group. For example, a group whose required chunks appear at ranks two and seven is complete at rank seven.

CERecall@$k$ is the fraction of questions with a complete evidence group within the first $k$ results. CEMRR@$k$ averages the reciprocal of the earliest completion rank, assigning zero when no group is complete within $k$. We report these metrics separately for T06 and T07, at rank 20 in Experiment 2 and at the selected retrieval cutoffs in Experiment 1.

\subsection{Latency and memory footprint measurements}\label{sec:costdesign}
\subsubsection{Retrieval latency in Experiment 1}
The three neural embedding models run on the same NVIDIA A100 80~GB PCIe GPU (Supplementary Table~\ref*{supp-tab:exp1system}). For each question, we time one retrieval call returning 1,000 source chunks. This includes query encoding, GPU-to-CPU vector transfer, Weaviate hybrid search, result receipt, and chunk mapping. Model loading, corpus and index preparation, output serialization, metric computation, and disk writes are excluded. Each question is timed once, without warm-up or repeated calls; we report mean latency overall and by question category.

\subsubsection{Component time in Experiment 2}
Experiment 2 uses an RTX 5070 Ti workstation for retrieval and a DGX Spark for BGE/Qwen reranking and Qwen3-4B-Instruct-2507 verse-fragment extraction (Supplementary Table~\ref*{supp-tab:exp2systems}). The FVR-off comparison records query embedding, EN retrieval, EN-to-chunk mapping, matched-budget direct retrieval, each branch's reranking, and CombMAX fusion. We sum these components per question, counting the shared embedding once, and then average across all questions or within each category. Each question has equal weight in the overall mean. This is an estimated serial computation time, not a measured end-to-end response time. These recorded timings cover branch scoring and merging before the additional top-$S_q$ truncation; the truncation was applied to saved rankings without rerunning retrieval or reranking. The component breakdown shows where the cost arises; it does not assume parallel execution. Timings describe one recorded execution per question and include the reranker-call wrapper overhead.

FVR extraction, lookup, and priority merging are recorded separately for the FVR-on comparison. Gemini has no comparable timing record. Supplementary Material, Section~\ref*{supp-app:evaluation-fields} lists the component fields. Experiment 1 and Experiment 2 use different hardware, retrieval depths, and timing scopes, and are reported separately.

\subsubsection{Memory footprint for chunk storage in Experiment 1}
We measure the \emph{memory footprint for chunk storage} as the disk space occupied by each Weaviate collection, with the same 60,121 source chunks per collection (Table~\ref{tab:storagefootprint}). Totals include chunk text, embeddings, indexes, and collection logs, and are reported in binary units (MiB and GiB).

\section{Results}\label{sec:results}
We organize the results around five comparisons: direct SV and MV retrieval against BM25; the contribution of EN; BGE versus Qwen reranking and its computation cost; FVR enabled versus disabled; and the combined pipelines versus Gemini. The component comparisons use the representation selected from the direct-retrieval comparison.

\subsection{BGE-M3 balances retrieval quality and resource use}\label{sec:representationresults}
\subsubsection{BGE-M3 leads in coverage; ColBERTv2 in early precision}
\begin{figure*}[!htbp]
\centering
\includegraphics[width=0.97\textwidth]{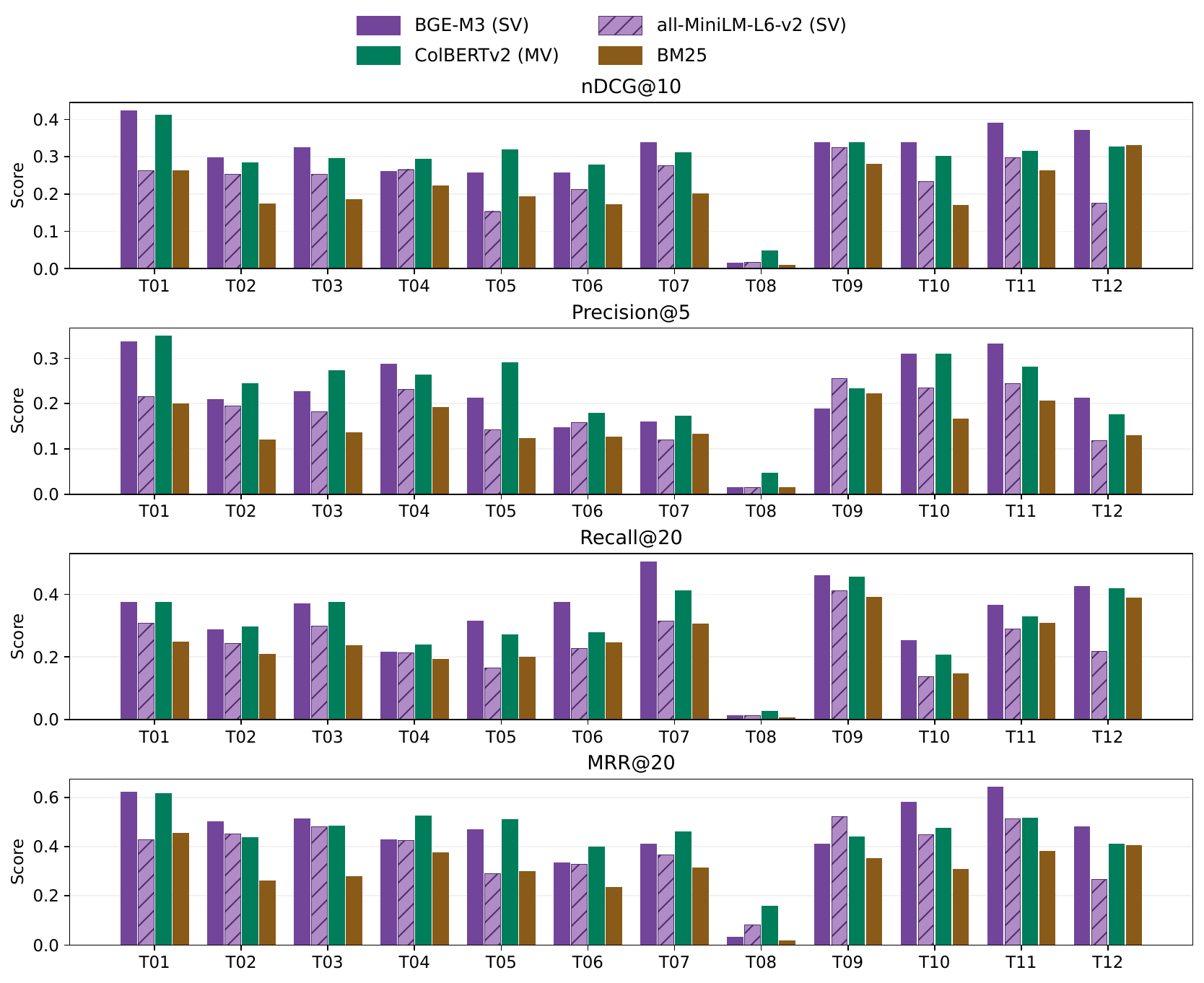}
\Description{Four vertical grouped-bar panels show nDCG at ten, precision at five, recall at twenty, and MRR at twenty across the twelve categories for BGE-M3, all-MiniLM-L6-v2, ColBERTv2, and BM25.}
\caption{Direct retrieval by question category: BGE-M3 and all-MiniLM-L6-v2 SV, ColBERTv2 MV, and BM25.}
\label{fig:representationcategories}
\end{figure*}

Figure~\ref{fig:representationcategories} compares BGE-M3 and all-MiniLM-L6-v2 SV retrieval, ColBERTv2 MV retrieval, and the BM25 baseline from \dhara{} Bench~\cite{dharaBenchmark2026}. The panels report nDCG@10, Precision@5, Recall@20, and MRR@20 across all twelve categories.

BGE-M3 exceeds BM25 in nDCG@10, Recall@20, and MRR@20 in every category, and in Precision@5 in ten. ColBERTv2 exceeds BM25 in Recall@20, Precision@5, and MRR@20 in every category; its nDCG@10 is higher in eleven, with T12 the exception (0.326 versus 0.331). These results establish the neural configurations' gains over the lexical baseline while retaining category-specific differences.

Comparing BGE-M3 with ColBERTv2, BGE-M3 has higher nDCG@10 and Recall@20 in eight categories. MRR@20 favors each model in six categories, while Precision@5 favors BGE-M3 in three categories and ColBERTv2 in eight, with a tie in T10. Both have nDCG@10 below 0.05 in Verse \& Textual Retrieval (T08), motivating a separate verse-matching component. Additional F1@10 and retrieval-depth plots appear in Supplementary Material, Section S1. Neither BGE-M3 nor ColBERTv2 leads across all category-metric combinations.

all-MiniLM-L6-v2 exceeds BM25 in nDCG@10, Precision@5, and MRR@20 in nine categories each, and in Recall@20 in seven. Its Recall@20 is lower than BGE-M3 and ColBERTv2 in eleven categories, while it exceeds BGE-M3 in T08. These results show the effectiveness trade-off associated with its smaller storage footprint and lower retrieval time.

\subsubsection{Deeper retrieval increases recall; F1 peaks at eight}
\begin{figure*}[!htbp]
\centering
\includegraphics[width=0.97\textwidth]{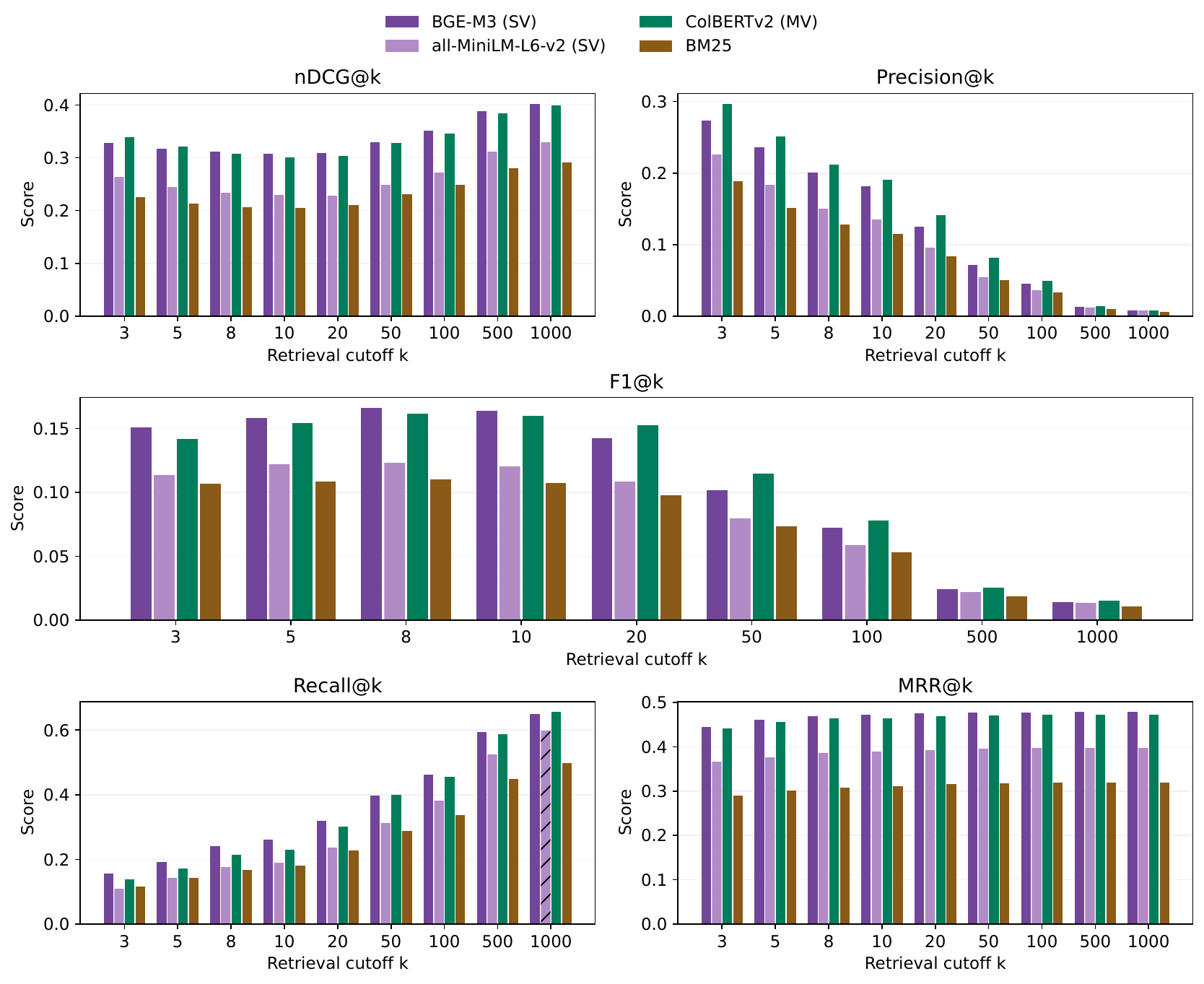}
\Description{Five grouped vertical-bar panels show overall nDCG, Precision, Recall, MRR, and F1 across nine retrieval cutoffs for BGE-M3, all-MiniLM-L6-v2, ColBERTv2, and BM25. Every bar averages 325 per-question scores.}
\caption{Experiment 1 overall scores across nine retrieval cutoffs; each bar averages all 325 questions.}
\label{fig:representationoverall}
\end{figure*}

Figure~\ref{fig:representationoverall} reports overall Precision, Recall, F1, nDCG, and MRR at $k\in\{3,5,8,10,20,50,100,500,1000\}$. Each bar averages the per-question score over all 325 questions in the \dhara{} Bench evaluation set. These overall scores weight each question equally; they are not unweighted averages of the twelve category means.

Increasing retrieval depth improves coverage. Between $k=20$ and $k=1000$, overall recall rises from 0.319 to 0.650 for BGE-M3, from 0.301 to 0.656 for ColBERTv2, from 0.235 to 0.596 for all-MiniLM-L6-v2, and from 0.229 to 0.498 for BM25. MRR changes much less over the same range: from 0.476 to 0.478 for BGE-M3 and from 0.469 to 0.472 for ColBERTv2. Precision falls as more lower-ranked chunks enter the cutoff; the F1 and nDCG panels show the corresponding balance between coverage and ranking quality. Among the evaluated cutoffs, all four configurations reach their highest mean F1 at $k=8$: 0.166 for BGE-M3, 0.161 for ColBERTv2, 0.123 for all-MiniLM-L6-v2, and 0.110 for BM25.

The detailed category-by-cutoff plots and direct complete-evidence plots are provided in Supplementary Material, Section S1. The main effect of greater retrieval depth is increased evidence coverage, with little change to the rank of the first relevant chunk.

\subsubsection{BGE-M3 and ColBERTv2 retrieve complementary evidence}
Table~\ref{tab:representationoverlap} shows how many relevant source chunks appear in both models' top-20 results and how many are found only by BGE-M3 or only by ColBERTv2. Both representations contribute unique relevant chunks. For T01, 115 relevant question-chunk pairs are shared, 68 are unique to BGE-M3, and 89 are unique to ColBERTv2. For T10, the corresponding counts are 55, 55, and 64. Thus similar scores need not mean that the same source chunks are retrieved.

\begin{table}[!htbp]
\caption{Relevant source chunks shared by BGE-M3 and ColBERTv2 or unique to each in the top-20 rankings, by category. Counts are question-chunk pairs.}
\label{tab:representationoverlap}
\centering\small
\begin{tabular}{@{}lrrr@{}}
\toprule
Category & Shared & BGE-M3 only & ColBERTv2 only \\
\midrule
T01 & 115 & 68 & 89 \\
T02 & 56 & 40 & 67 \\
T03 & 29 & 21 & 31 \\
T04 & 34 & 28 & 25 \\
T05 & 36 & 31 & 45 \\
T06 & 23 & 12 & 13 \\
T07 & 16 & 11 & 5 \\
T08 & 4 & 1 & 5 \\
T09 & 30 & 12 & 13 \\
T10 & 55 & 55 & 64 \\
T11 & 63 & 46 & 56 \\
T12 & 15 & 10 & 31 \\
\bottomrule
\end{tabular}

\end{table}

\subsubsection{BGE-M3 reduces retrieval time and storage relative to ColBERTv2}
\begin{figure}[!htbp]
\centering
\includegraphics[width=\linewidth]{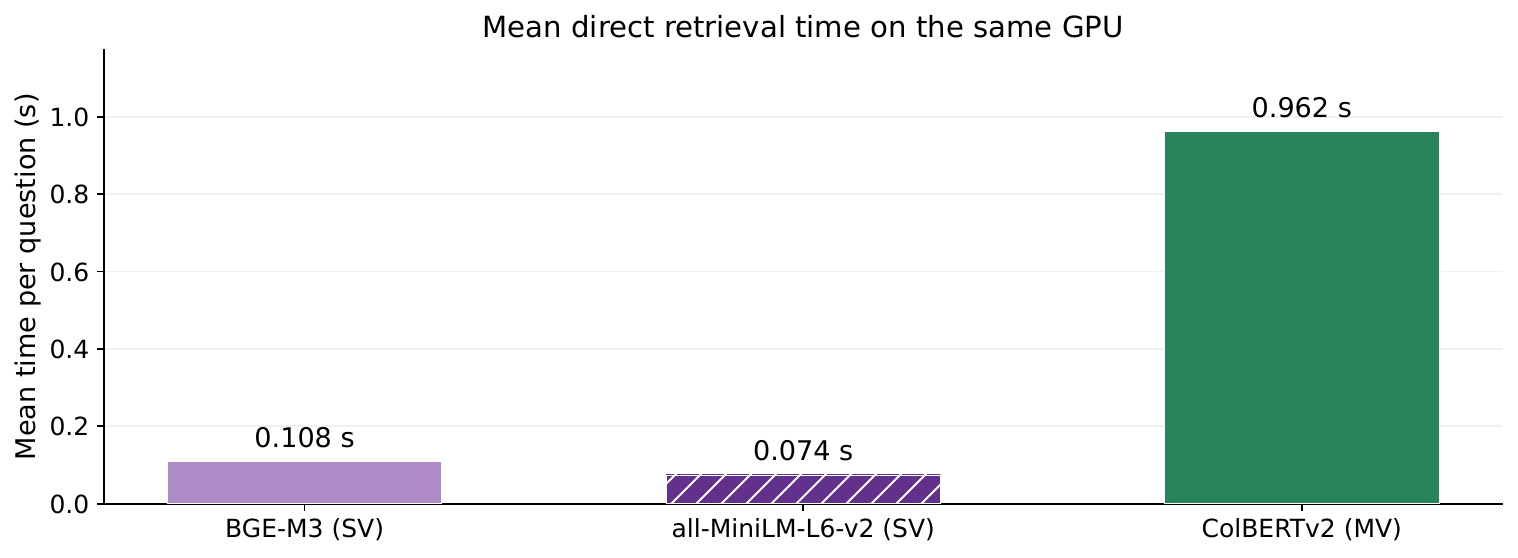}
\Description{Three vertical bars compare mean direct retrieval time for BGE-M3, all-MiniLM-L6-v2, and ColBERTv2 across all 325 questions, under the same hardware configuration and timed retrieval operations.}
\caption{Experiment 1 mean direct retrieval time across 325 questions, using the same hardware configuration and timing scope for all three models (Section~\ref{sec:costdesign}).}
\label{fig:representationoveralltiming}
\end{figure}

Figure~\ref{fig:representationoveralltiming} shows the mean retrieval time across all 325 questions, with all three embedding models evaluated under the same hardware configuration and using the same timed retrieval operations. BGE-M3 averages 0.108~s per question, ColBERTv2 0.962~s, and all-MiniLM-L6-v2 0.074~s. ColBERTv2's mean retrieval time per question is approximately 8.9 times that of BGE-M3.

Category mean times range from 0.099 to 0.122~s for BGE-M3, 0.556 to 1.751~s for ColBERTv2, and 0.065 to 0.086~s for all-MiniLM-L6-v2. Supplementary Material, Section S1, provides the timing bars by category. Table~\ref{tab:storagefootprint} compares collections containing the same 60,121 source chunks. ColBERTv2 has approximately 13.2 times the memory footprint for chunk storage of BGE-M3 (7.30~GiB versus 568.3~MiB). all-MiniLM-L6-v2 has the smallest memory footprint for chunk storage at 404.8~MiB, alongside lower F1@10 in every category and lower Recall@20 in eleven categories than the other two models.

\begin{table}[!htbp]
\caption{Memory footprints for chunk storage on disk: 60,121 source chunks per collection, including text, embeddings, indexes, and logs.}
\label{tab:storagefootprint}
\centering\small
\begin{tabular}{@{}llr@{}}
\toprule
Model & Representation & Memory footprint (disk) \\
\midrule
all-MiniLM-L6-v2 & Single vector & 404.8~MiB \\
BGE-M3 & Single vector & 568.3~MiB \\
ColBERTv2 & Multivector & 7.30~GiB \\
\bottomrule
\end{tabular}
\end{table}

BGE-M3 combines broadly comparable category scores with lower observed latency and memory footprint for chunk storage than ColBERTv2. Based on this balance of retrieval effectiveness, latency, and memory footprint for chunk storage, we select BGE-M3 as the base for Experiment 2. The smaller stored collection also supports this choice when planning for corpus expansion.

\subsection{EN improves retrieval at a matched retained candidate count}\label{sec:svresults}\label{sec:enresults}
\begin{figure*}[p]
\centering
\includegraphics[width=0.97\textwidth,height=.90\textheight,keepaspectratio]{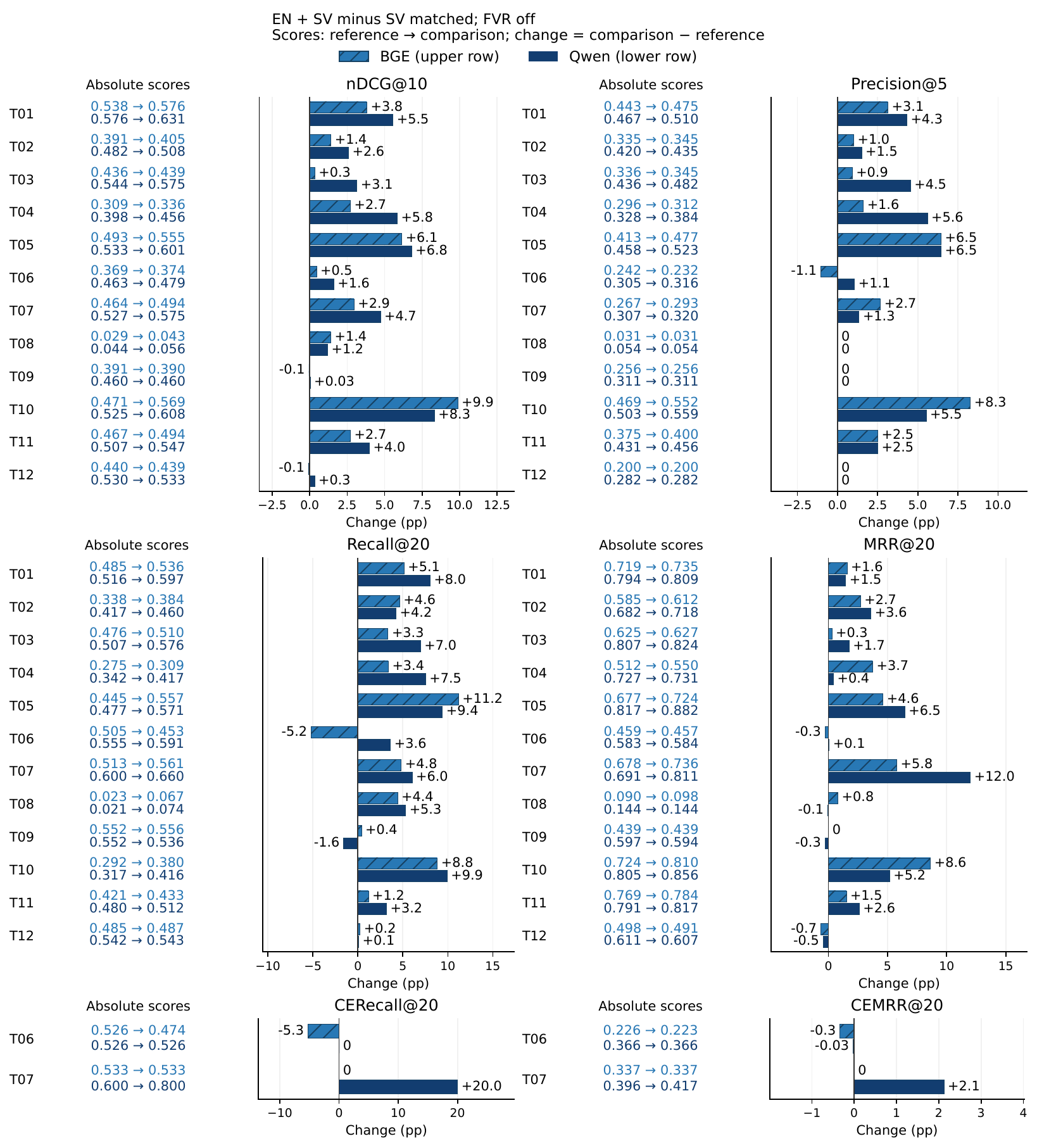}
\Description{Six horizontal difference panels compare adding EN to matched-budget SV for BGE and Qwen, including complete-evidence recall and MRR for T06 and T07.}
\caption{EN+SV minus SV matched, with FVR off. EN+SV and SV matched each retain $S_q$ candidates. Score pairs show SV matched to EN+SV; bars show changes in percentage points (pp). Bottom panels report complete evidence.}
\label{fig:svmatcheddifferences}
\end{figure*}

The component study uses BGE-M3 single-vector retrieval, denoted SV, based on the direct-retrieval results. Figure~\ref{fig:svmatcheddifferences} compares EN+SV against SV matched with the same reranker and FVR disabled. SV matched retrieves as many candidates as the EN branch; EN+SV then applies CombMAX to the two reranked lists. The combined ranking is truncated to $S_q$ chunks before evaluation, matching the direct branch's candidate count.

\subsubsection{EN improves nDCG and recall across most categories}
EN+SV has higher nDCG@10 than SV matched in ten categories with BGE and all twelve with Qwen. Recall@20 improves in eleven categories with each reranker. The clearest gains occur in T10, where nDCG@10 rises by 0.099 with BGE and 0.083 with Qwen, and Lexical Understanding (T05), with gains of 0.061 and 0.068. In Chronology (T04), the corresponding gains are 0.027 and 0.058. Qwen's T09 nDCG@10 gain is less than 0.001, so a positive category count alone does not describe the size of the benefit.

The effect on early relevance is less uniform. Precision@5 improves in eight categories with BGE and nine with Qwen; MRR@20 improves in nine with each. BGE loses Recall@20 in T06, while Qwen loses it in T09. Both rerankers have a small MRR@20 decrease in T12. Adding EN improves source discovery and ranking in many categories, but does not improve every metric.

\subsubsection{EN improves Multi-hop completeness with Qwen}
For Multi-hop (T07), EN+SV with Qwen retrieves a complete evidence group for 12 of 15 questions, compared with 9 of 15 for SV matched. CEMRR@20 also rises, from 0.396 to 0.417. With BGE, both conditions complete 8 of 15 questions and have CEMRR@20 of 0.337. For Composite (T06), Qwen remains at 10 of 19 complete questions, with a CEMRR@20 decrease below 0.001, while BGE decreases from 10 to 9 complete questions and from 0.226 to 0.223 CEMRR@20. EN's gains in individual-chunk metrics therefore do not consistently translate into more complete evidence groups.

\begin{figure*}[p]
\centering
\includegraphics[width=0.97\textwidth,height=.90\textheight,keepaspectratio]{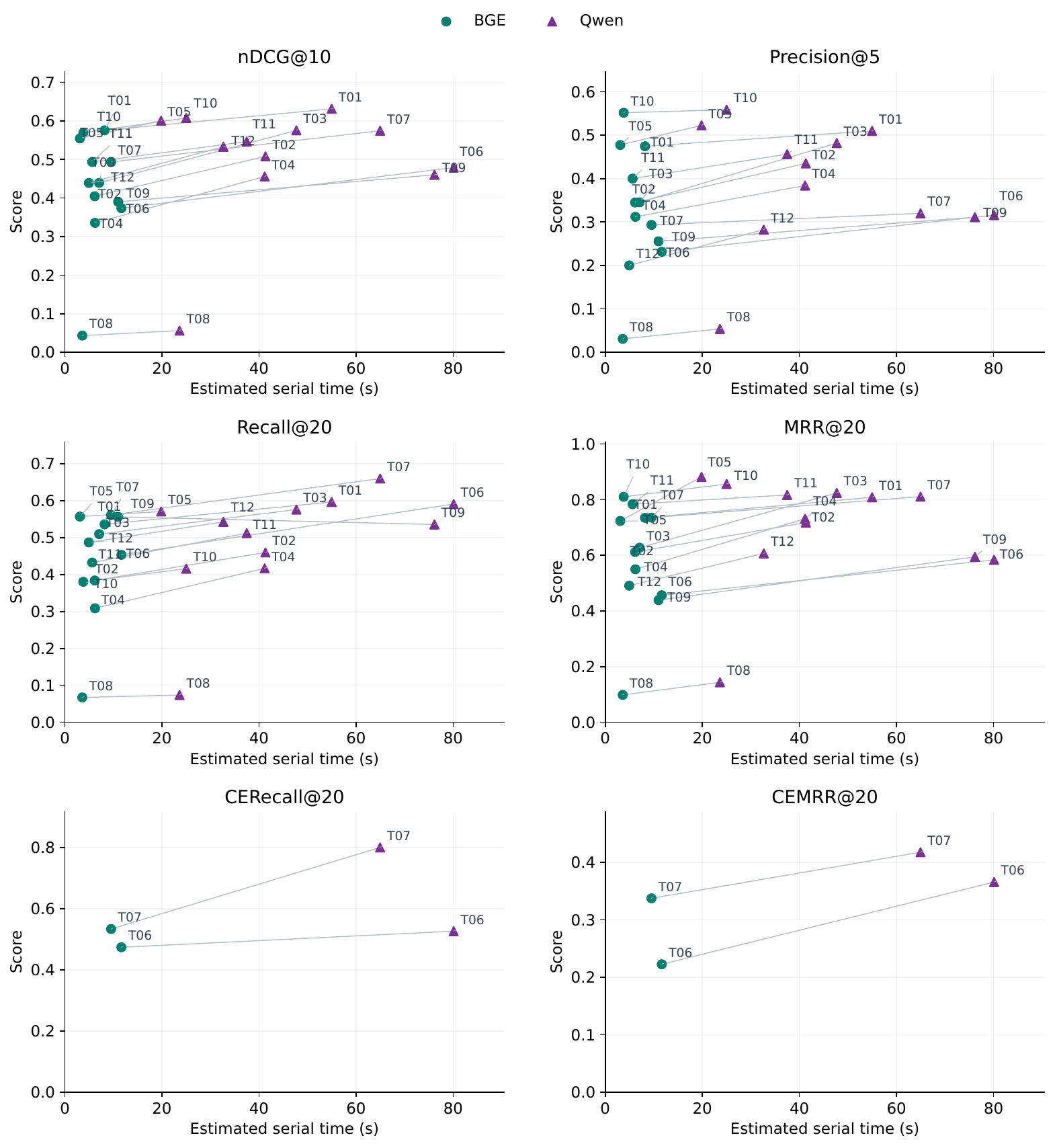}
\Description{Six scatter panels compare absolute category scores against estimated serial pipeline time for BGE and Qwen. Circles denote BGE and triangles denote Qwen; lines connect the same category. The bottom panels show complete-evidence measures for T06 and T07.}
\caption{BGE and Qwen retrieval scores versus estimated serial pipeline time for EN+SV, with FVR off. Lines connect the same question category; input pools and retained candidate counts are matched.}
\label{fig:rerankerdifferences}
\end{figure*}

\subsection{Qwen improves ranking quality at higher computation cost}\label{sec:svtiming}
\subsubsection{Qwen leads in nDCG, precision, and MRR across all categories}
Figure~\ref{fig:rerankerdifferences} compares Qwen with BGE using the same EN and direct input pools, retaining the top $S_q$ combined chunks for each reranker with FVR disabled. Each point shows a category's absolute score against its estimated serial pipeline time; lines connect the BGE and Qwen configurations for that category. Qwen has higher nDCG@10, Precision@5, and MRR@20 in all twelve categories. Recall@20 is higher in eleven; T09 is the exception, decreasing from 0.556 with BGE to 0.536 with Qwen. The largest nDCG@10 gains occur in T03 (0.136), T04 (0.120), and T06 (0.105).

Qwen also improves complete-evidence retrieval. In T06, CERecall@20 increases from 9/19 to 10/19 and CEMRR@20 from 0.223 to 0.366. In T07, CERecall@20 rises from 8/15 to 12/15 and CEMRR@20 from 0.337 to 0.417. Qwen's advantage thus extends beyond finding the first relevant chunk to retrieving complete supporting groups.

\subsubsection{Reranking dominates the computation cost}
Table~\ref{tab:svcategorytiming} reports mean component times across all 325 questions for the two configurations. The overall estimated serial time is 6.459~s with BGE and 43.412~s with Qwen, making Qwen approximately 6.72 times as costly in this timing scope. Across categories, the means range from 3.09 to 11.64~s with BGE and from 19.83 to 80.10~s with Qwen. Qwen costs 6.41-6.94 times as much in this timing scope. Composite (T06) has the highest category mean for both rerankers.

Across all questions, reranking the two branches accounts for 95.1\% of BGE's estimated serial time and 99.3\% of Qwen's. CombMAX fusion averages approximately 1~ms with either reranker. The time cost is therefore dominated by scoring the candidate chunks, rather than combining the ranked chunks.

Lexical Understanding (T05) illustrates the trade-off: Qwen raises nDCG@10 from 0.555 to 0.601 and MRR@20 from 0.724 to 0.882, while estimated serial time rises from 3.09 to 19.83~s. Qwen provides stronger ranking quality in this comparison; BGE provides a substantially lower computation cost.

\begin{table}[!htbp]
\centering
\caption{Mean component times across all 325 questions for EN+SV with FVR off. Each question has equal weight; units are shown per component.}
\label{tab:svcategorytiming}
\small
\setlength{\tabcolsep}{4pt}
\begin{tabular}{@{}llrr@{}}
\toprule
Component & Unit & BGE & Qwen \\
\midrule
Embedding & ms & 29.411 & 29.411 \\
EN retrieval & ms & 25.750 & 25.750 \\
EN mapping & ms & 19.323 & 19.323 \\
SV retrieval & ms & 241.616 & 241.616 \\
EN reranking & s & 2.806 & 19.700 \\
SV reranking & s & 3.336 & 23.395 \\
CombMAX fusion & ms & 0.948 & 1.074 \\
\midrule
Estimated serial total & s & 6.459 & 43.412 \\
\bottomrule
\end{tabular}

\end{table}

\subsection{FVR improves Verse-category retrieval with both rerankers}\label{sec:fvrresults}
\begin{figure}[!htbp]
\centering
\includegraphics[width=\linewidth]{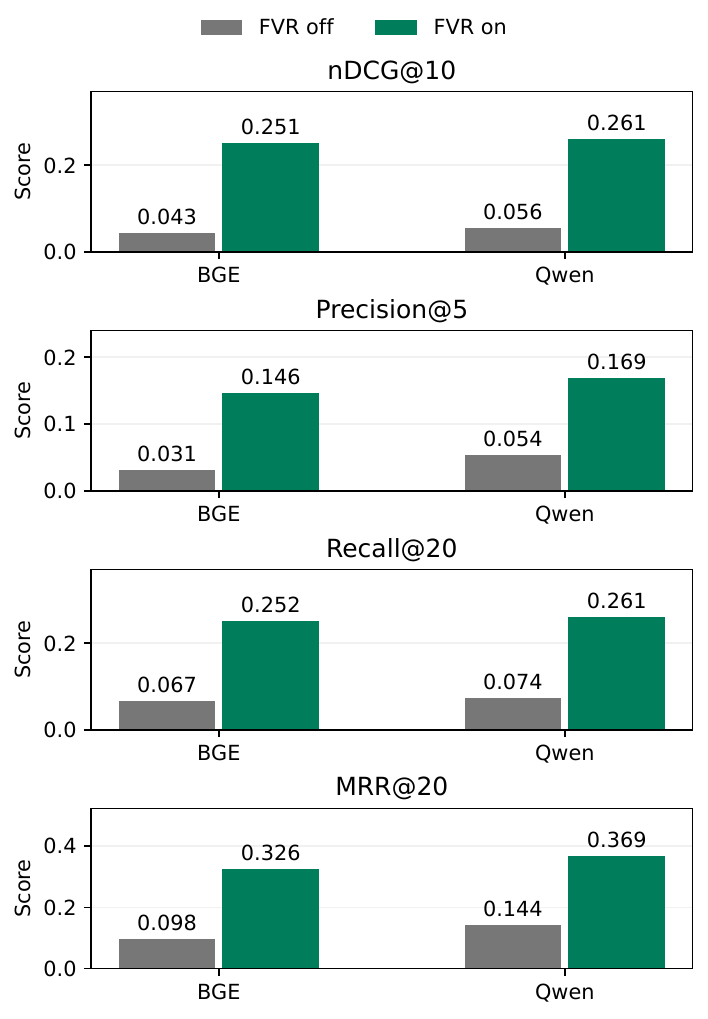}
\Description{Four grouped vertical-bar panels compare absolute T08 scores with FVR off and on for BGE and Qwen.}
\caption{FVR enabled and disabled in T08, holding the merged EN+SV ranking fixed before FVR.}
\label{fig:svfvronoff}
\end{figure}

Figure~\ref{fig:svfvronoff} compares FVR enabled and disabled in T08 while holding the merged EN+SV ranking fixed before FVR. With BGE, nDCG@10 rises from 0.043 to 0.251 and Recall@20 from 0.067 to 0.252. With Qwen, the corresponding changes are 0.056 to 0.261 and 0.074 to 0.261. FVR increases Precision@5 by 0.115 with both rerankers and raises MRR@20 by 0.228 with BGE and 0.225 with Qwen.

FVR changes the final ranking for 12 of the 325 questions with each reranker. Its gains are concentrated in T08, showing the value of matching verse fragments directly even after semantic retrieval and reranking. The complete-evidence scores in T06 and T07 are unchanged. Supplementary Material, Section S2 gives the full category comparison, including small decreases in T10 and T11 caused by the priority merge. FVR is most useful for fragment-bearing questions; its impact depends on the relevance of the chunks it places at the top.

\begin{figure*}[p]
\centering
\includegraphics[width=0.97\textwidth,height=.90\textheight,keepaspectratio]{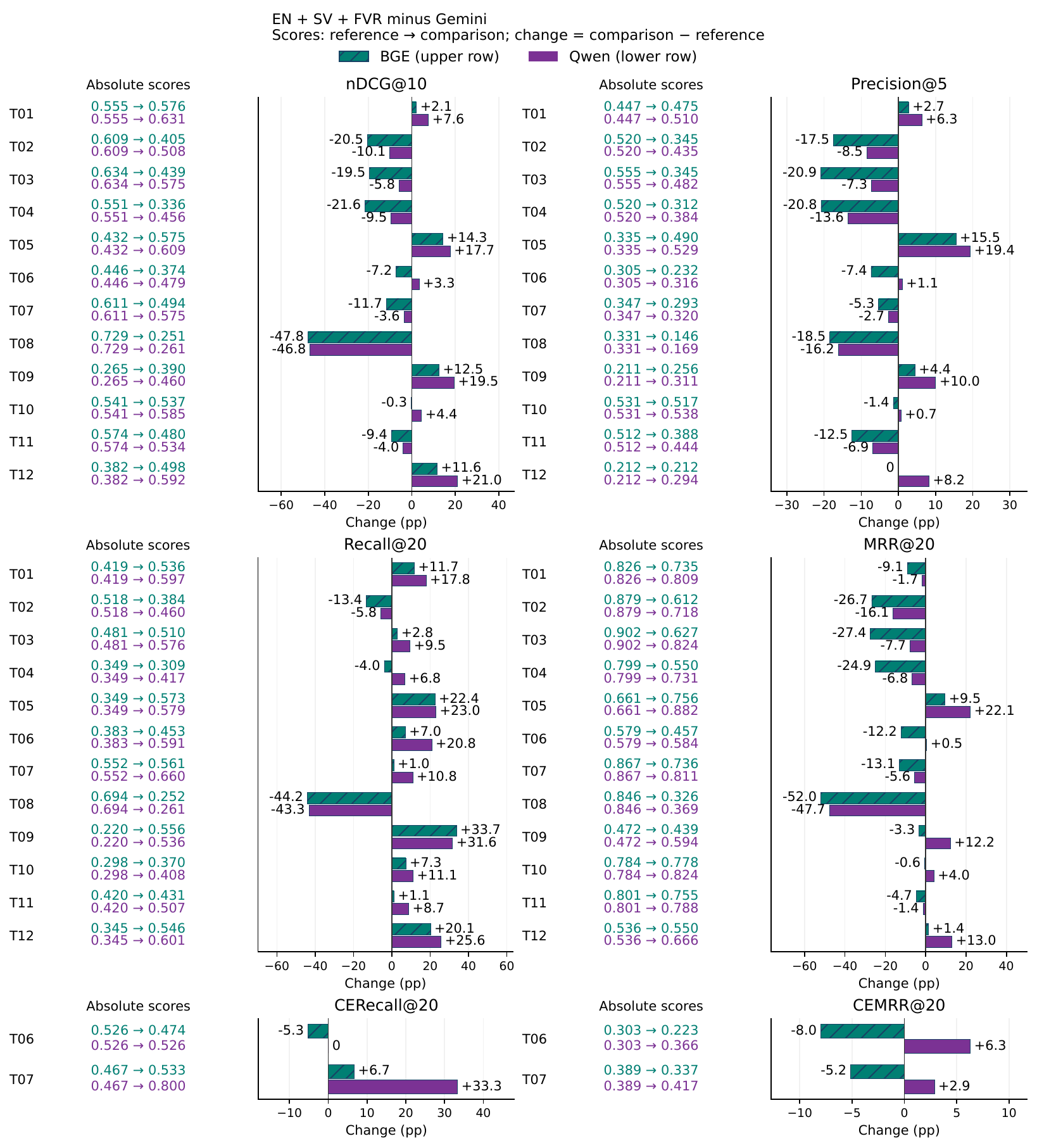}
\Description{Six horizontal difference panels compare both EN plus SV plus FVR pipelines with the same Gemini reference across categories.}
\caption{EN+SV with FVR versus Gemini, for BGE and Qwen. Score pairs show Gemini to EN+SV+FVR; bars show changes in percentage points (pp). Complete-evidence metrics apply to T06 and T07.}
\label{fig:geminidifferences}
\end{figure*}

\subsection{EN+SV offers broader coverage; Gemini leads in selected categories}\label{sec:geminiresults}
Figure~\ref{fig:geminidifferences} compares EN+SV+B+FVR and EN+SV+Q+FVR with Gemini, using the mapping and manual review procedure in Section~\ref{sec:geminidesign}. The same Gemini rankings serve as the reference for both rerankers.

\subsubsection{Coverage gains over Gemini do not imply earlier relevance}
The Qwen configuration has higher Recall@20 than Gemini in ten categories and higher nDCG@10 and Precision@5 in six each. MRR@20 is higher in five. The BGE configuration has higher Recall@20 in nine categories, nDCG@10 in four, Precision@5 in three, and MRR@20 in two. Thus the choice of reranker materially affects the comparison with Gemini.

EN+SV+Q+FVR exceeds Gemini on all four metrics in Lexical Understanding (T05), False Premise (T09), and T12. Gemini exceeds both pipelines on all four metrics in List / Set Retrieval (T02) and Verse \& Textual Retrieval (T08). In T08, Gemini's nDCG@10 is 0.729 and Recall@20 is 0.694, compared with 0.261 for both measures in the Qwen pipeline. FVR's improvement over the FVR-off condition does not eliminate this difference.

\subsubsection{Qwen improves complete-evidence ranking over Gemini}
For T06, Qwen and Gemini each complete 10 of 19 questions, but Qwen has higher CEMRR@20 (0.366 versus 0.303). BGE completes 9 questions, with CEMRR@20 of 0.223. For T07, Qwen completes 12 of 15 questions versus 7 for Gemini and has higher CEMRR@20 (0.417 versus 0.389). BGE completes 8 questions but has lower CEMRR@20 (0.337) than Gemini.

The combined pipeline is strongest relative to Gemini in evidence coverage, with Qwen also improving complete-evidence ranking. Gemini retains advantages in several categories and in the rank of the first relevant chunk. Supplementary Material, Section S2 provides absolute scores and the FVR-off comparison, allowing the EN+SV and verse contributions to be examined separately.

\section{Discussion}
The five comparisons separate decisions about representation, source discovery, reranking, verse matching, and the use of an LLM-based evidence retriever. Their practical implications depend on both the intended question categories and the amount of supporting evidence required.

\subsection{BGE-M3 provides a practical quality-cost balance}
The BM25 comparison establishes a useful starting point: BGE-M3 improves nDCG@10, Recall@20, and MRR@20 in every category, while ColBERTv2 also improves these measures in almost every category. Between BGE-M3 and ColBERTv2, neither model dominates all metrics. ColBERTv2 has higher Precision@5 in eight categories, while BGE-M3 has higher Recall@20 in eight; applications emphasizing the first few passages may therefore value the two representations differently. BGE-M3 combines broadly comparable category scores with a mean direct retrieval time of 0.108~s, compared with 0.962~s for ColBERTv2, under the same hardware configuration. Its memory footprint for chunk storage on disk is 568.3~MiB rather than 7.30~GiB for the same 60,121 chunks. ColBERTv2 therefore requires approximately 13.2 times the storage footprint and 8.9 times the retrieval time in this comparison. These measured costs favor BGE-M3 when planning for corpus growth under storage and response-time constraints; they do not establish how either system scales across corpus sizes.

The compact all-MiniLM-L6-v2 encoder reduces latency and chunk-storage footprint further, but its F1@10 is lower than both other neural models in every category. Choosing a single-vector architecture still requires choosing an encoder whose effectiveness is appropriate for the workload.

\subsection{Retrieval depth should follow the evidence requirement}
Increasing retrieval depth raises recall much more than it changes the rank of the first relevant chunk. BGE-M3's recall increases from 0.319 at rank 20 to 0.650 at rank 1,000; BM25 increases from 0.229 to 0.498. All four configurations reach their highest mean F1 at $k=8$ among the tested cutoffs, even though recall continues to increase at greater depths. This distinguishes the depth used to discover candidates from the number of passages ultimately shown or passed to a later stage. A larger discovery pool can improve coverage, while a short output ranking requires effective selection.

The shared and unique relevant chunks in Table~\ref{tab:representationoverlap} show that BGE-M3 and ColBERTv2 return partly different relevant evidence. Merging their rankings is therefore a candidate for improving coverage when the application can accommodate both indexes and retrieval paths. This would retain the larger MV storage footprint and add retrieval work; the benefit of the merged ranking would need to be evaluated against those costs.

\subsection{EN improves evidence discovery with one-time preprocessing and an additional reranking stage}
EN+SV improves nDCG@10 over SV matched in all categories with Qwen and ten with BGE. Recall@20 also improves in eleven categories with each reranker. However, Qwen's T09 nDCG@10 gain is below 0.001, and both rerankers slightly reduce MRR@20 in T12. EN is useful as a complementary discovery route, but its value should be judged by the size of the gain and the metric required by the task.

EN+SV and direct retrieval each retain $S_q$ candidates before evaluation. EN changes which chunks are selected, rather than increasing the retained candidate count. EN construction and indexing are performed once for unchanged source material, and the resulting notes and source links are reused across queries. At query time, adding EN requires note retrieval, source mapping, and one additional branch reranking stage alongside direct retrieval. CombMAX reuses the two branches' existing scores, with no further reranking of the merged list. The construction cost is thus shared across subsequent searches, while the additional branch work remains a per-query cost.

Complete evidence gives a more specific guide for Composite and Multi-hop workloads. Adding EN with Qwen improves T07 coverage from 9 to 12 complete questions out of 15, and also improves CEMRR@20. Qwen does not increase the number of complete T06 questions. BGE does not improve T07 complete-evidence scores and loses one complete T06 question. The benefit of EN depends on how the combined evidence is ranked, not only on how much is found.

\subsection{Qwen favors ranking quality; BGE favors lower computation cost}
On identical input candidate pools, with the combined rankings capped at $S_q$ and FVR disabled, Qwen improves nDCG@10, Precision@5, and MRR@20 in every category. Its overall estimated serial time is 43.412~s per question, compared with 6.459~s for BGE, a factor of 6.72 (Table~\ref{tab:svcategorytiming}). Reranking accounts for 99.3\% and 95.1\% of these totals, respectively. Deployments with sufficient GPU capacity and an acceptable response-time budget can favor Qwen for stronger ranking quality and complete-evidence retrieval. BGE is the less costly choice when computation time is the binding constraint. The measured factor applies to the evaluated deployment; GPU capacity alone does not determine the speedup on another system.

CombMAX fusion is inexpensive because it reuses the existing reranker scores. Reducing merge overhead would therefore have little effect on the measured total. Candidate selection and reranker execution are more consequential targets for reducing computation. Such changes would need to preserve the evidence gains observed here. T09 also illustrates why ranking and coverage must be checked separately: Qwen has higher nDCG@10 but lower Recall@20 than BGE.

\subsection{FVR strengthens retrieval for partially remembered verses}
FVR produces a clear T08 improvement when applied to fixed EN+SV rankings: approximately 20 percentage points in nDCG@10 and 19 percentage points in Recall@20 with either reranker. This supports a dedicated fragment-matching path for partially remembered verses, alongside the semantic and entity-based retrieval branches.

The final ranking changes for only 12 questions, so FVR is a targeted component. Priority insertion can also displace relevant chunks, as the T10 and T11 differences show. Fragment acceptance and merge order should be assessed together when adapting the method to another collection.

\subsection{Qwen narrows several gaps with Gemini and strengthens coverage}
Replacing BGE with Qwen makes the combined EN+SV+FVR pipeline more competitive with Gemini. The number of categories in which the pipeline exceeds Gemini rises from four to six for nDCG@10, from three to six for Precision@5, from nine to ten for Recall@20, and from two to five for MRR@20. Qwen therefore narrows several ranking deficits and produces stronger results in some categories, rather than uniformly matching Gemini. Gemini still has higher MRR@20 in seven categories. Broader evidence coverage and the earliest relevant passage are therefore distinct design objectives. Lexical Understanding, False Premise, and T12 favor the Qwen pipeline on all four conventional metrics; List / Set Retrieval and Verse \& Textual Retrieval favor Gemini. In T08, the remaining nDCG@10 gap is substantial: 0.261 for the Qwen pipeline versus 0.729 for Gemini, even after FVR is enabled.

Complete-evidence metrics sharpen this comparison. In T06, Qwen and Gemini complete the same number of questions, but Qwen has higher CEMRR@20. In T07, Qwen has higher complete-evidence coverage and CEMRR@20 than Gemini. For tasks requiring several supporting passages, these measures give a more relevant basis for choosing a configuration than the first relevant chunk alone. The results support evaluating candidate discovery, ranking quality, and evidence completeness together for the intended workload.

\section{Conclusion}
We presented \dhara{}, an IKS retrieval engine combining hybrid source-chunk search, Entity Notes (EN), Fuzzy Verse Retrieval (FVR), and neural reranking. Evaluation on the 325 questions in \dhara{} Bench shows how these components affect evidence discovery, ranking quality, and computation cost across question categories.

BGE-M3 exceeds the BM25 baseline in nDCG@10, Recall@20, and MRR@20 in every category. Its broadly comparable category scores to ColBERTv2, lower retrieval time, and smaller memory footprint for chunk storage on disk make it a practical base when planning for corpus growth. ColBERTv2 retains an early-precision advantage in several categories and retrieves some relevant chunks that BGE-M3 misses. Combining the two remains a direction for evaluation where the additional storage and retrieval costs are acceptable.

EN improves nDCG@10 in all twelve categories with Qwen and ten with BGE while retaining the same candidate count as direct retrieval. Notes and source links are prepared once for unchanged source material and reused across queries. The additional query-time work comprises EN retrieval, source mapping, and a branch reranking stage; CombMAX reuses the existing scores without reranking the merged list. Qwen achieves higher nDCG@10, Precision@5, and MRR@20 than BGE in every category, at approximately 6.72 times the overall estimated serial computation time. This makes the choice between the rerankers a trade-off between ranking quality and the available computation budget. FVR complements these components by improving Verse-category nDCG@10 by approximately 20 percentage points with either reranker.

The EN+SV+Qwen+FVR pipeline is more competitive with Gemini than its BGE counterpart, exceeding Gemini's Recall@20 in ten categories and improving both complete-evidence measures for Multi-hop questions. Gemini retains advantages in several categories and in early relevance. The results support combining entity-focused discovery, task-specific verse matching, and reranking according to the evidence required, while assessing coverage, complete-evidence retrieval, response-time requirements, and storage costs together.

\renewcommand{\acksname}{Acknowledgements}
\begin{acks}
% Organization names checked against the following public sources:
% DST SHRI: https://dst.gov.in/science-and-heritage-research-initiative-shri
% Perfios CSR policy: https://perfios.ai/wp-content/uploads/2025/04/csr_policy.pdf
% Adani legal name: https://www.adanienterprises.com/investors/corporate-governance
% IIT Madras project records identify Adani Enterprises Limited as the sponsor
% of the Adani Indology Research Scholarship (project CR24250952HSADAN008962).
% Adiyen company profile: https://in.linkedin.com/company/adiyen
An early proof of concept for this work was supported by the Department of Science and Technology (DST), Government of India, through its Science and Heritage Research Initiative (SHRI), under grant no.~DST/TDT/SHRI-34/2021. This work was supported by a Corporate Social Responsibility (CSR) grant from Perfios Software Solutions Private Limited to HTSR Institute. One of the joint first authors was supported through the Adani Indology Fellowship, funded by Adani Enterprises Limited.

We thank Adiyen Technologies Private Limited for assistance with data preparation.
\end{acks}

\FloatBarrier
\bibliographystyle{ACM-Reference-Format}
\bibliography{references}

\end{document}